\documentclass[sigconf]{acmart}
\AtBeginDocument{%
  \providecommand\BibTeX{{%
    \normalfont B\kern-0.5em{\scshape i\kern-0.25em b}\kern-0.8em\TeX}}}

\usepackage{amsmath}
\usepackage{bm}
\usepackage{enumitem}

\newtheorem{theorem}{\textbf{Theorem}}
\newtheorem{lemma}{\textbf{Lemma}}
\newtheorem{corollary}{\textbf{Corollary}}
\usepackage{algorithm,algorithmic}
\usepackage{multirow}
\usepackage{graphicx}
\usepackage{epstopdf}

\definecolor{myred}{rgb}{0.68627451, 0.14117647, 0.09803922}

\copyrightyear{2023}
\acmYear{2023}
\setcopyright{acmlicensed}\acmConference[WWW '23]{Proceedings of the
ACM Web Conference 2023}{May 1--5, 2023}{Austin, TX, USA}
\acmBooktitle{Proceedings of the ACM Web Conference 2023 (WWW '23),
May 1--5, 2023, Austin, TX, USA}
\acmPrice{15.00}
\acmDOI{10.1145/3543507.3583223}
\acmISBN{978-1-4503-9416-1/23/04}
\begin{document}

\title{On the Theories Behind Hard Negative Sampling for Recommendation}

\settopmatter{authorsperrow=4}

\author{Wentao Shi}
\email{shiwentao123@mail.ustc.edu.cn}
\affiliation{%
  \institution{University of Science and Technology of China}
  \city{Hefei}
  \country{China}
 }

\author{Jiawei Chen}
\email{sleepyhunt@zju.edu.cn}
\affiliation{%
  \institution{Zhejiang University}
  \city{Hangzhou}
  \country{China}
  }
\authornotemark[1]
  
\author{Fuli Feng}
\email{fulifeng93@gmail.com}
\affiliation{%
  \institution{University of Science and Technology of China}
  \city{Hefei}
  \country{China}
 }
  
\author{Jizhi Zhang}
\email{cdzhangjizhi@mail.ustc.edu.cn}
\affiliation{%
  \institution{University of Science and Technology of China}
  \city{Hefei}
  \country{China}
  }
  
\author{Junkang Wu}
\email{jkwu0909@gmail.com}
\affiliation{%
  \institution{University of Science and Technology of China}
  \city{Hefei}
  \country{China}
  }

\author{Chongming Gao}
\email{chongming.gao@gmail.com}
\affiliation{%
  \institution{University of Science and Technology of China}
  \city{Hefei}
  \country{China}
  }

\author{Xiangnan He}
\email{xiangnanhe@gmail.com}
\affiliation{%
  \institution{University of Science and Technology of China}
  \city{Hefei}
  \country{China}
  }
\authornote{*Corresponding author}

\renewcommand{\shortauthors}{Wentao et al.}
\begin{abstract}

    Negative sampling has been heavily used to train recommender models on large-scale data, wherein sampling hard examples usually not only accelerates the convergence but also improves the model accuracy. Nevertheless, the reasons for the effectiveness of Hard Negative Sampling (HNS) have not been revealed yet. In this work, we fill the research gap by conducting thorough theoretical analyses on HNS. Firstly, we prove that employing HNS on the Bayesian Personalized Ranking (BPR) learner is equivalent to optimizing One-way Partial AUC (OPAUC). Concretely, the BPR equipped with Dynamic Negative Sampling (DNS) is an exact estimator, while with softmax-based sampling is a soft estimator. Secondly, we prove that OPAUC has a stronger connection with Top-$K$ evaluation metrics than AUC and verify it with simulation experiments. These analyses establish the theoretical foundation of HNS in optimizing Top-$K$ recommendation performance for the first time. On these bases, we offer two insightful guidelines for effective usage of HNS: 1) the sampling hardness should be controllable, e.g., via pre-defined hyper-parameters, to adapt to different Top-$K$ metrics and datasets; 2) the smaller the $K$ we emphasize in Top-$K$ evaluation metrics, the harder the negative samples we should draw. Extensive experiments on three real-world benchmarks verify the two guidelines. 

\end{abstract}




\maketitle
\section{Introduction}\label{section1}


\begin{figure}[t]
  \centering
  \includegraphics[width=0.95\linewidth]{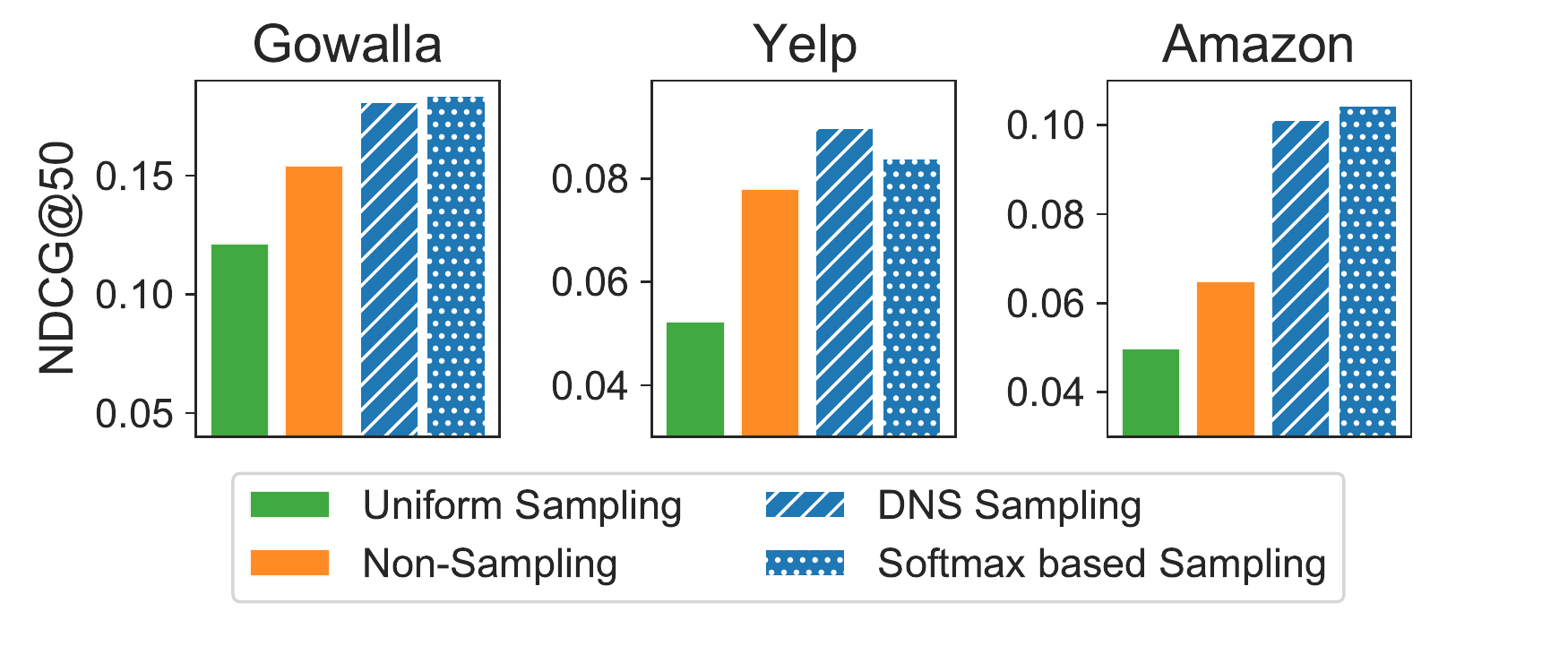}
  \caption{The recommendation performance on three widely used datasets with different sampling strategies.}
  \label{fig:Sampling}
  \Description{HNS strategies substantially outperform the Non-Sampling strategy. All of them outperform uniform sampling strategy.} 
\end{figure}

\begin{figure}[t]
  \centering
  \includegraphics[width=0.95\linewidth]{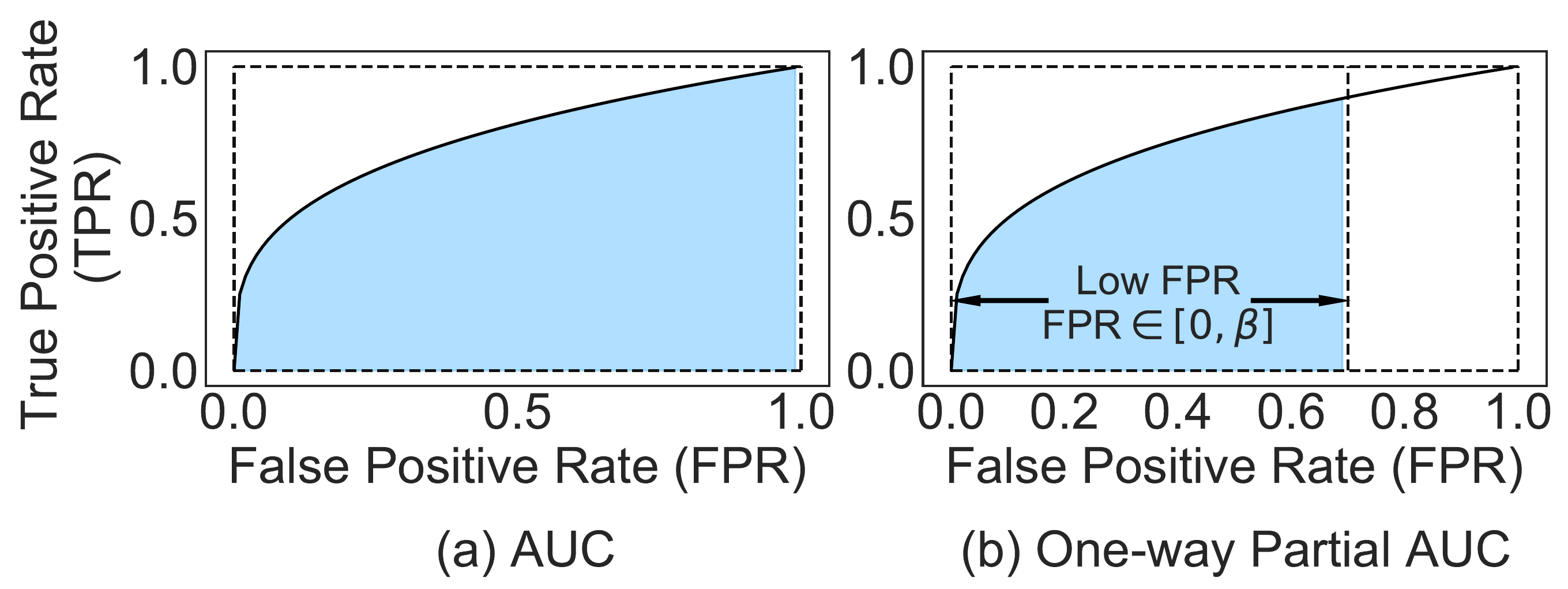}
  \caption{(a) AUC, measures the entire area of the ROC curve. (b) OPAUC, measures partial area within an FPR range of $[0,\beta]$. AUC is a special case of OPAUC with $\beta=1$.}
  \Description{Examples about AUC and partial AUC.}
  \label{fig:OPAUC}
\end{figure}

Recommendation systems are essential in addressing information overload by filtering unintended information and have benefited many high-tech companies \cite{DBLP:journals/corr/abs-2010-03240}. Bayesian Personalized Ranking (BPR) \cite{10.5555/1795114.1795167} is a common choice for learning recommender models from implicit feedback, which randomly draws negative items for the sake of efficiency and approximately optimizes the AUC metric. However, uniformly sampled negative items may not be informative, contributing little to the gradients and the convergence \cite{10.1145/2556195.2556248, 10.1145/2484028.2484126}. To overcome this obstacle, researchers have proposed many Hard Negative Sampling (HNS) methods, such as Dynamic Negative Sampling (DNS) \cite{10.1145/2484028.2484126} and Softmax-based Sampling methods \cite{10.1145/3077136.3080786, 10.1145/3366423.3380187, 10.1145/3485447.3512075}. Superior to uniform sampling, HNS methods oversample high-scored negative items, which are more informative with large gradients and thus accelerate the convergence \cite{10.1145/3450289}. 

\begin{figure}[t]
  \centering
  \includegraphics[width=0.95\linewidth]{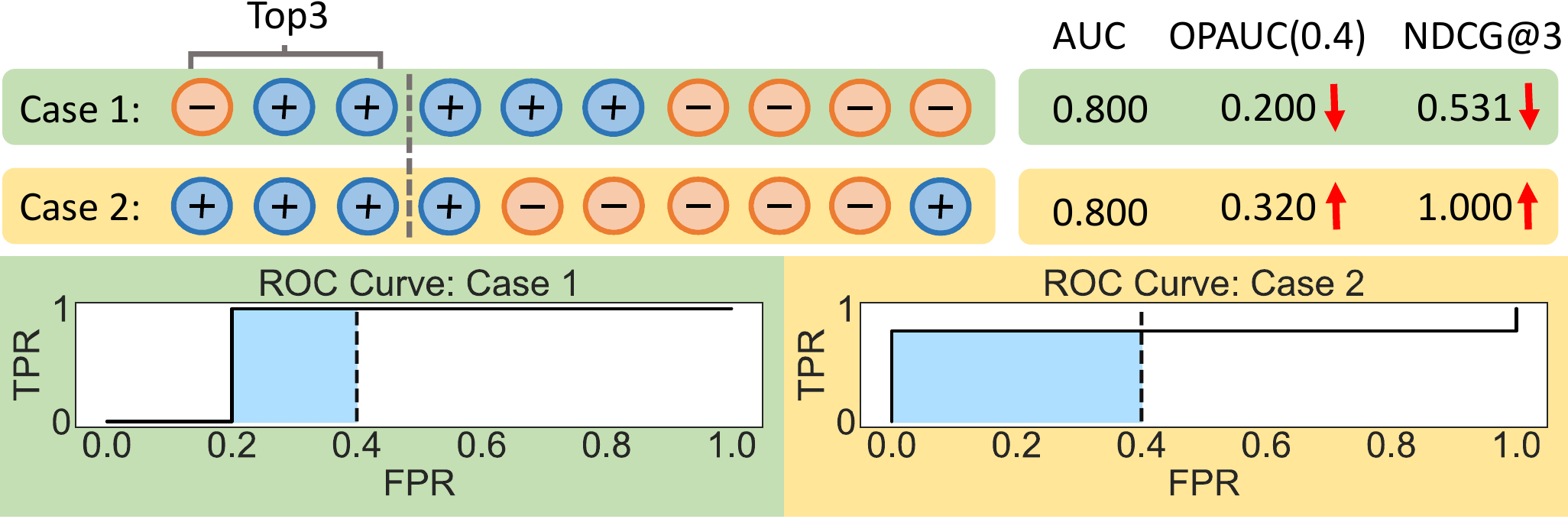}  \caption{Two simple cases have the same overall ranking performance but different top-ranking performance. The ROC curves of two cases have the same AUC but different OPAUC($\beta$=0.4).}
  \label{fig:case}
  \Description{Two simple cases show OPAUC has stronger connection with Top-$K$ metrics.}
  \vspace{-0.4cm}
\end{figure}

While existing work usually attributes the superior performance of HNS to its better convergence, we find that the merits of HNS are beyond this thought. 
To validate it, we conduct empirical analyses on three widely used datasets in Figure \ref{fig:Sampling}. We compare two HNS strategies with a strong baseline named Non-Sampling\footnote{All compared methods optimize the same loss of BPR.} \cite{10.1145/3522672} that computes the gradient over the whole data (including all negative items). As such, the Non-Sampling strategy is supposed to converge to a better optimum more stably \cite{10.1145/3373807, chen2020efficient, DBLP:conf/sigir/HeZKC16, DBLP:conf/www/BayerHKR17}.
Nevertheless, to our surprise, both HNS strategies substantially outperform the Non-Sampling strategy. It indicates that fast convergence may not be the only justification for the impressive performance of HNS. There must be other reasons for its superior performance, which motivates us to delve into HNS and explore its theoretical foundation. Our findings are twofold:

\begin{itemize}[leftmargin=*]
    \item \textbf{Optimizing the BPR loss equipped with HNS is equivalent to optimizing the One-way Partial AUC (OPAUC)}, whereas the original BPR loss only optimizes AUC. OPAUC puts a restriction on the range of false positive rate (FPR) $\in [0,\beta]$ \cite{10.2307/3695437}, as shown in Figure~\ref{fig:OPAUC}(b), which emphasizes the ranking of top-ranked negative items. In contrast, AUC is a special case of OPAUC($\beta$) with $\beta=1$, which considers the whole ranking list. Our proof of the equivalence is based on the \textit{Distributionally Robust Optimization} (DRO) framework \cite{1908.05659} (\textit{cf.} Section~\ref{section3}).
    \item \textbf{Compared to AUC, OPAUC has a stronger connection with Top-$K$ metrics.} To illustrate it, we conduct simulation studies with randomly generated ranking lists, showing that OPAUC exhibits a much higher correlation with Top-$K$ metrics like Recall, Precision and NDCG by tuning $\beta$~(\textit{cf.} Figure~\ref{OPAUC_simulation}). This is because both OPAUC and Top-$K$ metrics care more about the ranking of top-ranked items, as shown in Figure \ref{fig:case}. Furthermore, we confirm the correlation through theoretical analysis that Recall@$K$ and Precision@$K$ metrics could be higher and lower bounded with a function of specific OPAUC($\beta$), respectively.
\end{itemize}  

\begin{figure}[t]
  \centering
  \includegraphics[width=0.95\linewidth]{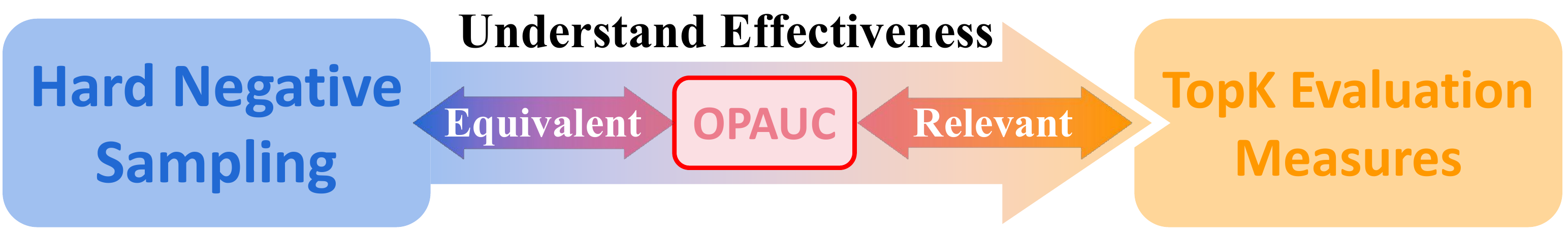}
  \caption{The relationship among HNS, OPAUC measure, and Top-$K$ evaluation metrics.}
  \Description{Optimizing the BPR loss equipped with HNS is equivalent to optimizing OPAUC. Compared to AUC, OPAUC has a stronger connection with Top-$K$ metrics.}
  \label{fig:relationship1}
\end{figure}

In short, our analyses reveal that equipping BPR with HNS is equivalent to optimizing the OPAUC, leading to better Top-$K$ recommendation performance (\textit{cf.} Figure~\ref{fig:relationship1}).
Our analyses not only explain the impressive performance of HNS but also shed light on how to perform HNS in recommendation. 
Given the correspondence between Top-$K$ evaluation metrics and OPAUC($\beta$), we offer two instructive guidelines to ensure the practical effectiveness of HNS. 
First, the sampling hardness should be controllable, e.g., via pre-defined hyper-parameters, to adapt to different Top-$K$ metrics and datasets. 
Second, the smaller the $K$ we emphasize in Top-$K$ evaluation metrics, the harder the negative samples we should draw.

The main contributions of this paper are summarized as follows:
\begin{itemize}[leftmargin=*]
    \item 
    We are the first to establish the theoretical foundations for HNS: equipping BPR with DNS is an exact estimator of the OPAUC objective, and with softmax-based sampling is a soft estimator.
    \item We conduct theoretical analyses, simulation studies, and real-world experiments, to justify the connection between OPAUC and Top-$K$ metrics and explain the performance gain of HNS.
    \item We provide two crucial guidelines on how to perform HNS and adjust sampling hardness. The experiments on real-world datasets validate the rationality of the guidelines.
\end{itemize}

\section{Background}\label{section2}

This section provides the necessary background of Implicit Feedback, Hard Negative Sampling Strategies, One-way Partial Area Under ROC Curve (OPAUC), and Distributionally Robust Optimization (DRO) \cite{1908.05659}. DRO is a robust learning framework that we will use in subsequent sections.
\subsection{Implicit Feedback}
The goal of a recommender is to learn a score function
\begin{math}
r(c,i|\mathbf{\theta})
\end{math} to predict scores of unobserved item $i$ in context $c$ and recommend the top-ranked items \cite{DBLP:conf/www/BayerHKR17}. A larger predicted score reflects a higher preference for the item $i \in \mathcal{I}$ in a context $c \in \mathcal{C}$\footnote{In collaborative filtering setting, a context $c$ denotes a user $u \in \mathcal{U}$; in sequential recommendation setting, $c$ can be a historical interaction sequence.}. In the implicit feedback setting, we can only observe positive class $\mathcal{I}_c^+ \subseteq \mathcal{I}$ in the context $c$. The remaining $\mathcal{I}_c^- = \mathcal{I}\backslash\mathcal{I}_c^+$ are usually considered as negative items in the context $c$. In personalized ranking algorithms with BPR loss, the objective functions can be formulated as follows:
\begin{equation}
\min_{{\theta}} \sum_{c\in \mathcal{C}} \sum_{i \in \mathcal{I}_c^+} E_{j \sim P_{ns}(j|c)} \left[ \ell \left( r(c,i|{\theta}) -r(c,j|{\theta}) \right) \right],
\label{implict_object}
\end{equation}
where $\mathbf{\theta}$ are the model parameters, $\ell(t)$ is the loss function which is often defined as $\log(1+\exp(-t))$. $P_{ns}(j|c)$ denotes the negative sampling probability that a negative item $j\in \mathcal{I}_c^-$ in the context c is drawn. In BPR \cite{10.5555/1795114.1795167}, each negative
item is assigned an equal sampling probability. For HNS strategies, a negative item with a larger predicted score will have a higher sampling probability. 
For ease of understanding, we refer to \cite{NEURIPS2020_0c7119e3} and \textbf{define the ``hardness'' of a negative sample as its predicted score, i.e., a negative sample is ``harder'' than another when its score is larger.}
In what follows, $r(c,i|\theta)$ is abbreviated as $r_{ci}$ for short.

\subsection{Hard Negative Sampling Strategies}
Different from static sampling like uniform and popularity-aware strategy \cite{10.1145/3097983.3098202}, HNS strategies are adaptive both to context and recommender models during the training. Here we review two widely-used HNS strategies. 

DNS \cite{10.1145/2484028.2484126} ranks the negative items and oversamples the high-ranked items\footnote{Without loss of generality, we consider a special case of DNS (Algorithm 2 in \cite{10.1145/2484028.2484126}) that set $n$ to $|\mathcal{I}_c^-|$, set $\beta_1, \cdots, \beta_{M-1}$ to 1 and set $ \beta_{M}, \cdots, \beta_{N}$ to 0. Our analysis can generalize to the arbitrary multi-nomial distribution of $\beta_k$.}. 
The sampling probability of DNS is defined as:
\begin{equation}
    P_{ns}^{DNS}(j|c) = 
    \begin{cases}
    \frac{1}{M}, & j \in S^{\downarrow}_{\mathcal{I}_c^-}[1, M] \\
    0, & j \in others \\
    \end{cases},
    \label{defi:DNS}
\end{equation}
where $S^{\downarrow}_{\mathcal{I}_c^-}[1, M] \subset \mathcal{I}_c^-$ denotes the subset of the top-ranked $M$ negative items, i.e., the negative samples with top-$M$ largest predicted scores. \textbf{Remark that the smaller the $M$ is, the harder the negative samples will be drawn.}\par
Softmax-based sampling is widely used in adversarial learning \cite{10.1145/3308558.3313416, 10.1145/3077136.3080786} and importance sampling \cite{10.1145/3366423.3380187, 10.1145/3485447.3512075}, where they refer to softmax distribution to assign higher sampling probability to higher scored items. The negative sampling probability can be defined as: 
\begin{equation}
    \begin{split}
    P_{ns}^{Softmax}(j|c) & = \frac{\exp(r_{cj}/\tau)}{\sum_{k\in \mathcal{I}_c^-} \ \exp(r_{ck}/\tau)} \\
    & = \frac{\exp((r_{cj}-r_{ci})/\tau)}{\sum_{k\in \mathcal{I}_c^-}\ \exp((r_{ck}-r_{ci})/\tau)},
    \end{split}
    \label{defi:Softmax}
\end{equation}
where $\tau$ is a temperature parameter. \textbf{It is noteworthy that the smaller the $\tau$ is, the harder the samples will be drawn.}

\subsection{One-way Partial AUC}
For each context $c$, we can define true positive rates (TPR) and false positive rates (FPR) as
\begin{equation}
    TPR_{c,\theta}(t) = \mathbf{Pr}(r_{ci}>t|i\in \mathcal{I}_c^+),
\end{equation}
\begin{equation}
    FPR_{c,\theta}(t) = \mathbf{Pr}(r_{cj}>t|j\in \mathcal{I}_c^-).
\end{equation}
Then, for a given $s\in[0,1]$, let $TPR_{c,\theta}^{-1}(s) = \inf \{ t\in \mathbb{R}, TPR_{c,\theta}(t)<s\}$ and $FPR_{c,\theta}^{-1}(s) = \inf \{ t\in \mathbb{R}, FPR_{c,\theta}(t)<s\}$. Based on these, the AUC can be formulated as 
\begin{equation}
    \text{AUC}(\theta) =\frac{1}{|\mathcal{C}|} \sum_{c\in\mathcal{C}} \int_0^1 TPR_{c,\theta}\left[FPR_{c,\theta}^{-1}(s) \right]\mathrm{d}s.
\end{equation}
As shown in Figure \ref{fig:OPAUC}, One-way Partial AUC (OPAUC) only cares about the performance within a given  false positive rate (FPR) range $[\alpha, \beta]$. Non-normalized OPAUC \cite{10.2307/3695437} is equal to 
\begin{equation}
    OPAUC(\theta, \alpha, \beta) = \frac{1}{|\mathcal{C}|} \sum_{c\in\mathcal{C}}\int_{\alpha}^{\beta} TPR_{c,\theta}\left[FPR_{c,\theta}^{-1}(s) \right]\mathrm{d}s.
\label{OPAUC}
\end{equation}
In this paper, we consider the special case of OPAUC with $\alpha=0$, which is denoted as $OPAUC(\beta)$ for short. Based on the definition in Eq. \eqref{OPAUC}, we can have the following non-parametric estimator of OPAUC($\beta$):
\begin{equation}
    \widehat{OPAUC(\beta)} = \frac{1}{|\mathcal{C}|} \sum_{c\in\mathcal{C}} \frac{1}{n_+} \frac{1}{n_-} \sum_{i\in\mathcal{I}_c^+} \sum_{j \in S^{\downarrow}_{\mathcal{I}_c^-}[1, n_-\cdot\beta]} \mathbb{I}(r_{ci}>r_{cj}),
    \label{non_para_OPAUC}
\end{equation}
where $n_+ = |\mathcal{I}_c^+|$ and $n_-=|\mathcal{I}_c^-|$, and $\mathbb{I}(\cdot)$ is an indicator function. For simplicity, we assume $n_-\cdot\beta$ is a positive integer.\par
Since the OPAUC estimator in Eq. \eqref{non_para_OPAUC} is non-continuous and non-differentiable, we usually replace the indicator function with a continuous surrogate loss $L(c,i,j) = \ell(r_{ci}-r_{cj})$. With suitable surrogate loss $\ell(\cdot)$, maximizing $\widehat{OPAUC(\beta)}$ in Eq. \eqref{non_para_OPAUC} is equivalent to the following problem: 
\begin{equation}
    \min_{\theta} \frac{1}{|\mathcal{C}|} \sum_{c\in\mathcal{C}} \frac{1}{n_+} \sum_{i\in\mathcal{I}_c^+}  \frac{1}{n_-\cdot\beta} \sum_{j \in S^{\downarrow}_{\mathcal{I}_c^-}[1, n_-\cdot\beta]} L(c,i,j).
    \label{object_OPAUC}
\end{equation}
Remark that the objective is divided by a fixed constant $\beta$ for proof, which does not affect the properties of the objective function. For surrogate loss $\ell(\cdot)$, \cite{10.5555/2832249.2832379} proposes a sufficient condition to ensure it consistent for OPAUC maximization, where the widely used logistic loss $\ell(t)=\log(1+\exp(-t))$ satisfies the properties. \par
Additionally, for comparison among different $\beta$, we define normalized OPAUC($\beta$) following \cite{article_Analy},
\begin{equation}
    \text{OPAUC}_{\text{norm}}(\beta) = \text{Trans}\left(\text{OPAUC}(\beta)\right),
    \label{noramlized_OPAUC}
\end{equation}

where the normalized transformation is defined as:
\begin{equation}
    \text{Trans}(A) = \frac{1}{2} \left[ 1+ \frac{A- \min_\theta A}{\max_\theta A - \min_\theta A} \right].
\end{equation}

\subsection{Distributionally Robust Optimization}
Given a divergence $D_\phi$ between two distributions $P$ and $Q$, Distributionally Robust Optimization (DRO) aims to minimize the expected risk over the worst-case distribution $Q$ \cite{1908.05659, 2155-3289_2022_1_159, NEURIPS2020_64986d86}, where $Q$ is in a divergence ball around training distribution $P$. Formally, it can be defined as:
\begin{equation}
    \begin{gathered}
        \min_{\theta} \sup_Q E_Q\left[ \mathcal{L}(f_\theta(\mathbf{x}), y) \right] \\
        s.t. \  D_\phi(Q||P) \leq \rho,
    \end{gathered}
    \label{equation:DRO}
\end{equation}
where the hyperparameter $\rho$ modulates the distributional shift, $\mathcal{L}$ is the loss function. In this paper, we will focus on two special divergence metrics, i.e. the KL divergence $D_{KL}(Q||P)=\int \log(\frac{\mathrm{d}Q}{\mathrm{d}P}) \mathrm{d}Q$ \cite{Hu2012KullbackLeiblerDC} and the CVaR divergence $D_{CVaR}(Q||P) = \sup \log(\frac{\mathrm{d}Q}{\mathrm{d}P})$ \cite{1810.08750}.

\section{Hard negative sampling meets OPAUC}\label{section3}

In this section, we prove that the BPR loss equipped with HNS optimizes OPAUC($\beta$), which is the first step to understanding the effectiveness of HNS. \par

\begin{figure}[t]
  \centering
  \includegraphics[width=0.95\linewidth]{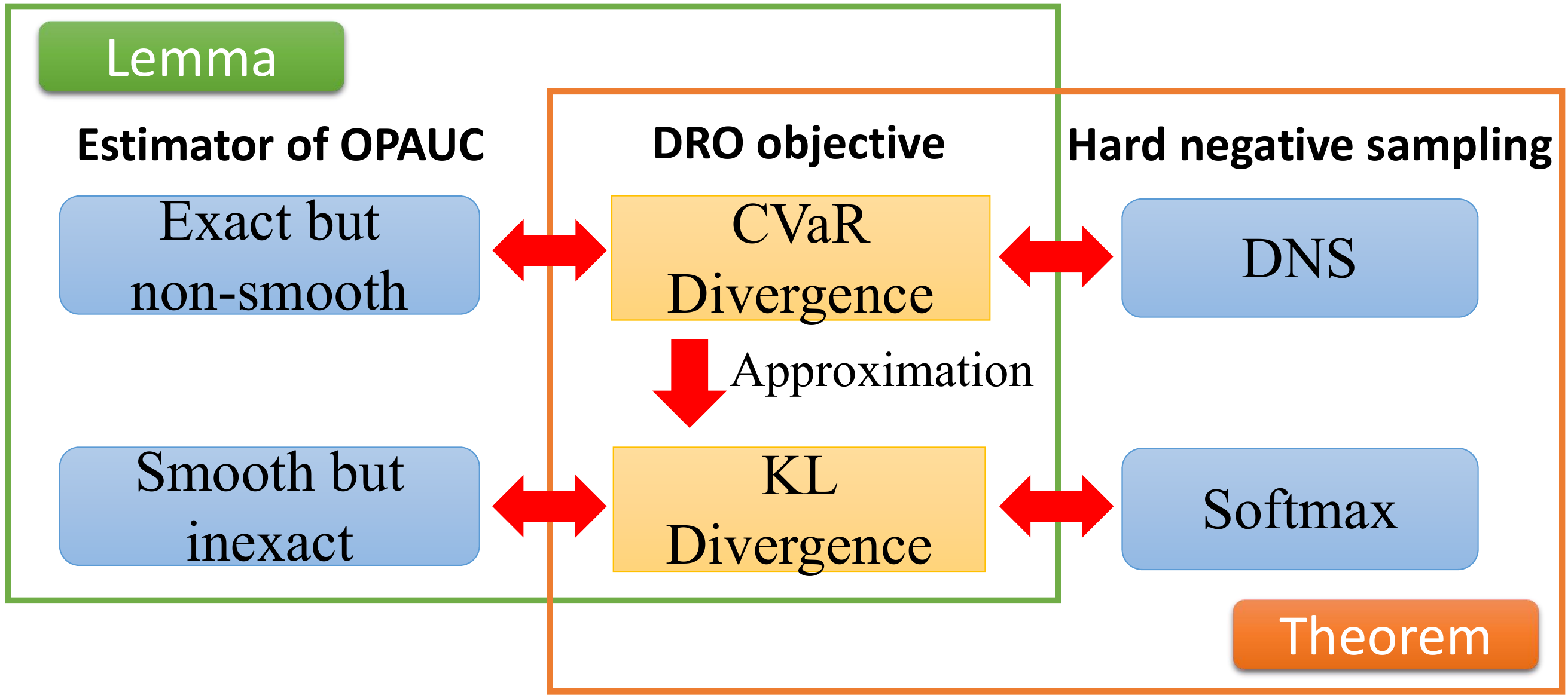}
  \caption{The Lemma \ref{lemma_CVAR_OPAUC} shows the equivalence between the OPAUC estimator and DRO objective. Based on DRO objective, we prove the equivalence between HNS and OPAUC in Theorem \ref{theorem_DNS_OPAUC} and Theorem \ref{Theorem_Softmax_OPAUC}.}
  \label{fig:OPAUC_DRO_hard}
  \Description{The proof outline of Theorem \ref{theorem_DNS_OPAUC} and Theorem \ref{Theorem_Softmax_OPAUC}.}
\end{figure}

We achieve the proof based on the DRO objective and present the proof outline in Figure \ref{fig:OPAUC_DRO_hard}. Following the theorems proposed in \cite{pmlr-v162-zhu22g}, we first show the connection between the OPAUC objective and the DRO-based objective. Then we prove that the personalized ranking problem (Eq. \eqref{implict_object}) equipped with HNS is equivalent to the DRO-based objective in our theorems. \par 

Following \cite{pmlr-v162-zhu22g}, we define the DRO-based objective as:
\begin{equation}
    \begin{gathered}
    \min_{\theta} \frac{1}{|\mathcal{C}|} \sum_{c\in\mathcal{C}}\frac{1}{n_+} \sum_{i\in\mathcal{I}_c^+} \max_{Q} E_{Q} \left[ L(c,i,j) \right]\\
    s.t. \ D_\phi(Q||P_0) \leq \rho,
    \end{gathered}
    \label{DRO_object}
\end{equation}
where $P_0$ denotes uniform distribution over $\mathcal{I}_c^-$, the hyperparameter $\rho$ modulates the degree of distributional shift, $D_\phi$ is the divergence measure between distributions. \par
Then we show the connection between the OPAUC objective and the DRO-based objective through the following lemma:
\begin{lemma}
[Theorem 1 of \cite{pmlr-v162-zhu22g}]By choosing CVaR divergence $D_\phi=D_{CVaR}(Q||P_0)=\sup \log(\frac{\mathrm{d}Q}{\mathrm{d}P_0})$ and setting $\beta=e^{-\rho}$, the DRO-based objective (Eq. \eqref{DRO_object}) is equivalent to the $OPUAC(\beta)$ objective (Eq. \eqref{object_OPAUC}).
\label{lemma_CVAR_OPAUC}
\end{lemma}
Based on the above lemma, we prove the equivalence between the OPAUC objective and the HNS based objective.

\begin{theorem}
By choosing $P_{ns}=P_{ns}^{DNS}$, 
\begin{equation}
    M=n_-\cdot\beta ,
\end{equation}
the DNS based problem (Eq. \eqref{implict_object}) is equivalent to the $OPUAC(\beta)$ objective (Eq. \eqref{object_OPAUC}).
\label{theorem_DNS_OPAUC}
\end{theorem}

\begin{figure*}[t]
  \centering
  \includegraphics[width=0.9\textwidth]{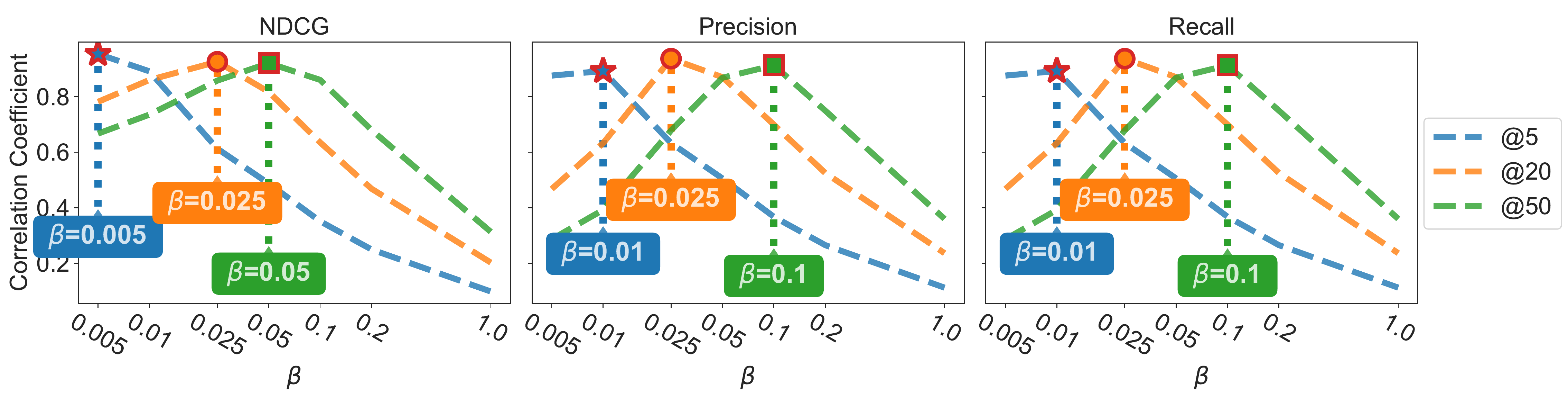}
  \caption{The estimated correlation coefficient between Top-$K$ evaluation metrics and OPAUC$_{norm}$($\beta$) under Monte Carlo sampling experiments, where $N_+=200$ and $N_-=800$. We highlight the value of $\beta$ when each curve reaches its maximum correlation coefficient. Remark that AUC is also a special case of OPAUC$_{norm}$($\beta$) with $\beta=1$.}
  \label{OPAUC_simulation}
  \Description{The estimated correlation coefficient between Top-$K$ evaluation metrics and OPAUC$_{norm}$($\beta$) under Monte Carlo sampling experiments.}
\end{figure*}

\begin{proof}
Given Lemma \ref{lemma_CVAR_OPAUC}, we just need to show that DNS sampling based problem (Eq. \eqref{implict_object}) is equivalent to the DRO-based objective (Eq. \eqref{DRO_object}).\par
By choosing CVaR divergence, then the DRO-based objective (Eq. \eqref{DRO_object}) reduces to \cite{pmlr-v162-zhu22g} (using strong duality and Theorem 4 in \cite{rockafellar2017risk})
\begin{equation} 
\min_{\theta} \min_{\mathbf{\eta}\ge 0} \frac{1}{|\mathcal{C}|} \sum_{c\in\mathcal{C}} \frac{1}{n_+} \sum_{i\in\mathcal{I}_c^+} \{\frac{1}{e^{-\rho}} \cdot E_{j\sim P_0} \left[ \left(L(c,i,j) - \eta_i\right)_+ \right] + \eta_i  \},
\label{DNS_proof}
\end{equation}
where $P_0$ denotes uniform distribution over $\mathcal{I}_c^-$. Following \cite{pmlr-v139-zhai21a}, it's easy to see that the optimal $\eta_i$ is the $e^{-\rho}$-quantile of $L(c,i,j)$, which is defined as: 
\begin{equation}
    \eta_i^* = \inf_{\eta_i}\{ P_{j\sim P_0}[L(c,i,j)>\eta_i] <  e^{-\rho}\}.
\end{equation}
Substitute $\eta_i$ with $\eta_i^*$ in Eq. \eqref{DNS_proof} and replace $e^{-\rho}$ with $\frac{M}{n_-}$, then we obtain the equivalence between DNS sampling based problem (Eq. \eqref{implict_object}) and DRO-based objective (Eq. \eqref{DRO_object}). Recall the conclusion in Lemma \ref{lemma_CVAR_OPAUC}, then we complete the proof by setting $M=n_-\cdot \beta$.
\end{proof}

\textbf{Remark}: The DNS based problem is an exact but non-smooth estimator of OPUAC($\beta$), which is consistent for OPAUC($\beta$) maximization. \textbf{The hyperparameter $M$ in DNS strategy directly determines $\beta$ in the OPAUC objective}.

\begin{theorem}
By choosing $P_{ns}=P_{ns}^{Softmax}$, \begin{equation}
    \tau = \sqrt{\frac{\mathrm{Var}_j(L(c,i,j))}{-2\log\beta}},
    \label{tau_proof}
\end{equation}
\begin{equation}
     \mathrm{Var}_j(L(c,i,j)) = E_{j\sim P_0}\left[ (L(c,i,j) - E_{j\sim P_0}[L(c,i,j)])^2\right],
     \label{var}
\end{equation}
then problem (Eq. \eqref{implict_object}) equipped with softmax-based sampling strategy is a surrogate version of the $OPUAC(\beta)$ objective (Eq. \eqref{object_OPAUC}).
\label{Theorem_Softmax_OPAUC}
\end{theorem}

The proof process is similar to Theorem \ref{theorem_DNS_OPAUC}. Substitute CVaR divergence with KL divergence but remain the same $\rho$, then we get a soft estimator of OPAUC($\beta$). We prove that the soft estimator is equivalent to softmax-based sampling problem (Eq. \eqref{implict_object}). The precise relationship between $\tau$ and $\beta$ is complex and hard to compute. Hence we get an approximate version via the Taylor expansion. The detailed proof can be found in Appendix \ref{proof_of_softmax}.\par
\textbf{Remark}: The BPR loss equipped with softmax-based sampling is a smooth but inexact estimator of OPAUC($\beta$). \textbf{The hyperparameter $\tau$ in softmax-based sampling directly determines $\beta$ in OPAUC objective.}

\section{OPAUC meets Top-K metrics}\label{section4}

In this section, we investigate the connection between OPAUC($\beta$) and Top-$K$ evaluation metrics, which is the second step to understanding the effectiveness of HNS. We propose two arguments to declare their relationship:

\begin{enumerate}
    \item \textbf{Compared to AUC, OPAUC($\beta$) has a stronger correlation with Top-$K$ evaluation metrics.}
    \item \textbf{A smaller $K$ in Top-$K$ evaluation metrics has a stronger correlation with a smaller $\beta$ in OPAUC($\beta$).} 
\end{enumerate}
 We conduct theoretical analysis and simulation experiments to verify our proposals as follows.
\subsection{Theoretical Analysis}
In this subsection, we analyze the connection between OPAUC($\beta$) and Top-$K$ metrics from a theoretical perspective. To be concrete, we prove that given $K$, Precision@$K$ and Recall@$K$ are higher bounded and lower bounded by the functions of specific OPAUC($\beta$).
\begin{theorem}
Suppose there are $N_+$ positive items and $N_-$ negative items, where $N_+>K$ and $N_->K$. For any permutation of all items in descending order, we have
\begin{equation}
    \begin{split}
    \frac{1}{N_+}\left\lfloor \frac{N_++K-\sqrt{(N_++K)^2-4N_+N_-\times OPAUC(\beta)}}{2} \right\rfloor \\ \leq 
    Recall@K \leq \frac{1}{N_+} \left\lceil \sqrt{N_+N_-\times OPAUC(\beta)}\right\rceil,
    \end{split}
    \label{proof_recall}
\end{equation}
\begin{equation}
    \begin{split}
    \frac{1}{K}\left\lfloor \frac{N_++K-\sqrt{(N_++K)^2-4N_+N_-\times OPAUC(\beta)}}{2} \right\rfloor \\ \leq 
    Precision@K \leq \frac{1}{K} \left\lceil \sqrt{N_+N_-\times OPAUC(\beta)}\right\rceil,
    \end{split}
    \label{proof_precision}
\end{equation}
where $\beta=\frac{K}{N_-}$.
\label{theorem_topk_measure}
\end{theorem}

\textbf{Remark:} From above, we get the following inspirations: 
\begin{enumerate}[leftmargin=*]
    \item The Top-$K$ metrics like Precision@$K$ and Recall@$K$ have a strong connection with specific OPAUC($\beta$), where $\beta=\frac{K}{N_-}$. However, such a connection does not exist for AUC, which confirms our first argument. Hence, maximizing specific OPAUC($\beta$) approximately optimizes specific Precision@$K$ and Recall@$K$.
    \item The smaller the $K$ is, the smaller the $\beta$ ($=\frac{K}{N_-}$) should be considered. A smaller $K$ has a stronger connection with a smaller $\beta$, which effectively verifies our second argument.
\end{enumerate}

\subsection{Simulation Experiments}

In this subsection, we conduct Monte Carlo sampling experiments to analyze the connection between OPAUC($\beta$) and Top-$K$ evaluation metrics. For comparison among different $\beta$, we use normalized OPAUC defined in Eq. \eqref{noramlized_OPAUC} here. Suppose there are $N_+$ positive items and $N_-$ negative items in the item set $\mathcal{I}$. Due to the vast scale of the entire permutation space of items, it is impossible to enumerate all cases for analyses directly. Hence, we make a Monte-Carlo approximation and uniformly sample permutations from the space as simulated ranking lists 100000 times. Then we calculate the evaluation metrics (Top-$K$ metrics and OPAUC$_{norm}$($\beta$)) for these simulated ranking lists. Afterward, we estimate the correlation coefficient between Top-$K$ metrics and OPAUC$_{norm}$($\beta$) and report them in Figure \ref{OPAUC_simulation}. We report $\beta$ of OPAUC$_{norm}$($\beta$) in logarithmic scale. Furthermore, we highlight the value of $\beta$ when each curve reaches its maximum correlation coefficient. Remark that OPAUC$_{norm}$(1) is equal to AUC.
\par
As shown in Figure \ref{OPAUC_simulation}, we have the following observations: 
\begin{enumerate}[leftmargin=*]
    \item The correlation coefficient of the highest point of the curve is much larger than the correlation coefficient when $\beta$ is equal to 1. That means most Top-$K$ evaluation metrics have higher correlation coefficients with specific OPAUC$_{norm}$($\beta$) (above 0.8) than AUC (under 0.4), which clearly verifies our first argument.
    \item Given a specific $K$ in Top-$K$ metrics, the correlation coefficient with OPAUC$_{norm}$($\beta$) gets the maximum value at a specific $\beta$. Both too large and too small $\beta$ will degrade the correlation with specific Top-$K$ metrics.
    \item For different $K$, the peak of the curve varies according to $\beta$. The smaller the $K$ in the Top-$K$ evaluation metrics, the smaller the $\beta$ that takes the maximum value of the correlation coefficient. This effectively confirms our second argument.
    \item On the left side of the peak of the curve, we find that the correlation coefficient of NDCG@$K$ descends more slowly than the other two metrics. This is because NDCG@$K$ pays more attention to top-ranked items in Top-$K$ items.
\end{enumerate}

\section{Deep understanding of HNS}\label{section5}

Based on the arguments discussed above, we gain a deeper theoretical understanding of HNS. The BPR loss equipped with HNS optimizes OPAUC($\beta$), which has a stronger connection with Top-$K$ metrics. In this sense, we derive the following corollary: 
\begin{corollary}
    The BPR loss equipped with HNS approximately optimizes Top-$K$ evaluation metrics, where the level of sampling hardness controls the value of $K$.
\end{corollary}
Moreover, we take a step further and propose two instructive guidelines for effective usage of HNS.
\begin{enumerate}[leftmargin=*]
    \item \textbf{The sampling hardness should be controllable, e.g., via pre-defined hyper-parameters, to adapt to different Top-$K$ metrics and datasets.}
    \item \textbf{The smaller the $K$ we emphasize in Top-$K$ evaluation metrics, the harder the negative samples we should draw.}
\end{enumerate}
Motivated by these, we generalize the DNS and softmax-based sampling to two controllable algorithms DNS($M$, $N$) and Softmax-v($\rho$, $N$), as shown in Algorithm \ref{alg:DNS} and Algorithm \ref{alg:softmax} respectively. 
\begin{itemize}[leftmargin=*]
    \item In DNS($M$, $N$), we utilize hyperparameter $M$ to control sampling hardness, where the original DNS is a special case with $M=1$.
    \item In Softmax-v($\rho$, $N$), we propose to use an adaptive $\tau$ in Eq. \eqref{tau_proof}, instead of a fixed $\tau$ in Eq. \eqref{defi:Softmax}. Hyperparameter $\rho$ controls the sampling hardness. This ensures that $\beta$ of the optimization objective $OPAUC(\beta)$ remains the same during training. 
\end{itemize}

\begin{algorithm}[t]{\vspace*{-0.0in}}
    \centering
    \caption{DNS ($M$, $N$)}
    \label{alg:DNS}
    \begin{algorithmic}[1]  
    \STATE Initialize $\theta$
    \FOR {$t = 1,\ldots, T$}
        \STATE Sample a mini-batch $\mathcal{B} \in \mathcal{D}$
        \FOR{$(c,i) \in \mathcal{B}$}
       
         \STATE Uniformly sample a mini-batch $\mathcal{B}_c'\in\mathcal{I}_c^-$, $|\mathcal{B}_c'| =N$.
         
        \STATE Let $p_{cij}=\begin{cases}\frac{1}{M}, & j\in\mathcal{S}_{\mathcal{B}_c'}^{\downarrow}[1,M] \\
        0, &j \in others.
        \end{cases}$
        \ENDFOR
        \STATE Compute a gradient estimator $\nabla_t$ by 
        $$ \nabla_t = \frac{1}{|\mathcal{B}|} \sum_{(c,i)\in\mathcal{B}} \sum_{j\in\mathcal{I}_c^-} p_{cij} \nabla_\theta L(c,i,j).$$ 
        
        \STATE Update $\theta_{t+1} = \theta_t - \eta \nabla_t$.
    \ENDFOR
    \end{algorithmic}
\end{algorithm}

\begin{algorithm}[t]{\vspace*{-0.0in}}
    \centering
    \caption{Softmax-v ($\rho$, $N$)}
    \label{alg:softmax}
    \begin{algorithmic}[1]  
    \STATE Initialize $\theta$
    \FOR {$t = 1,\ldots, T$}
        \STATE Sample a mini-batch $\mathcal{B} \in \mathcal{D}$
        \FOR{$(c,i) \in \mathcal{B}$}
       
         \STATE Uniformly sample a mini-batch $\mathcal{B}_c'\in\mathcal{I}_c^-$, $|\mathcal{B}_c'| =N$.
         
        \STATE Let $p_{cij}=\begin{cases} \frac{e^{\ell(r_{ci}-r_{cj})/\tau}}{\sum_{k\in \mathcal{B}_c'}\ e^{\ell(r_{ci}-r_{ck})/\tau}}, &j\in\mathcal{B}_c' \\
        0, &j \in others,
        \end{cases}$ \\
        where $\tau=\sqrt{\frac{\mathrm{Var}_j(L(c,i,j))}{2\rho}}$.
        \ENDFOR
        \STATE Compute a gradient estimator $\nabla_t$ by 
        $$ \nabla_t = \frac{1}{|\mathcal{B}|} \sum_{(c,i)\in\mathcal{B}} \sum_{j\in\mathcal{I}_c^-} p_{cij} \nabla_\theta L(c,i,j).$$ 
        
        \STATE Update $\theta_{t+1} = \theta_t - \eta \nabla_t$.
    \ENDFOR
    \end{algorithmic}
\end{algorithm}

\begin{figure}[t]
  \centering
  \includegraphics[width=0.95\linewidth]{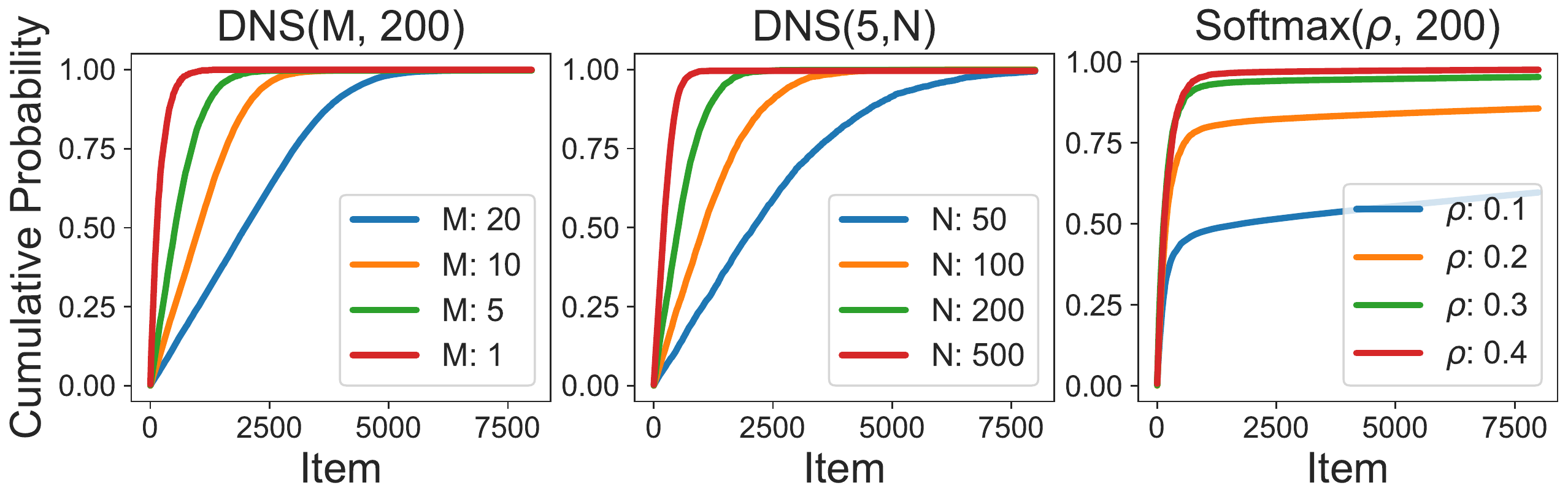}
  \caption{Approximated distributions under different parameter settings. The faster the curve rises, the higher the sampling probability the top-ranked items are drawn with.}
  \label{Cumulative_Probability}
  \Description{Approximated sampling distributions under different parameter settings.}
\end{figure}

As discussed, the hyperparameters $M$ and $\rho$ affect how hard the negative samples we will draw. Besides, the size of the sampling pool $N$ also affects the actual sampling probability of negative items. We conduct simulation experiments to investigate the difference of the sampling distribution under different parameter settings. We choose the user embeddings and items embeddings from the well-trained model on the Gowalla dataset and keep them fixed. Then, we randomly pick a (user, positive item) pair ($c$, $i$) and then simulate the sampling process 10000 times to estimate the actual sampling probability. The average value of $p_{cij}$ over the sampling process is approximated as the actual sampling probability that negative item $j$ is chosen by pair ($c$, $i$) for training. We report the cumulative probability distribution under different parameter settings in Figure \ref{Cumulative_Probability}. The negative items are in descending order w.r.t. their scores. \par

Since items are in descending order, we conclude that the faster the curve rises, the higher the sampling probability the top-ranked items are drawn with. Easily, we have the following observations:
\begin{itemize}[leftmargin=*]
    \item Smaller $M$ in DNS($M$, $N$) means higher sampling hardness.
    \item Larger $N$ in DNS($M$, $N$) means higher sampling hardness.
    \item Larger $\rho$ in Softmax-v($\rho$, N) means higher sampling hardness.
\end{itemize}

\section{Experiments}\label{section6}

In this section, we evaluate the models on three public datasets to figure out the following questions: 
\begin{itemize}[leftmargin=*]
    \item \textbf{(Q1)} How do DNS($M$, $N$) and Softmax-v($\rho$, $N$) perform compared to state-of-the-art HNS methods? Is it beneficial to control sampling hardness with pre-defined hyperparameters? 
    \item \textbf{(Q2)} Can experiment results validate our second guideline on adjusting sampling hardness according to $K$ in Top-$K$ metrics?
\end{itemize}

\paragraph{Dataset}
The Statistics of three public datasets are shown in Table \ref{tab:dataset}, which vary in scale and sparsity. The Gowalla dataset is the collection of user check-in histories. The Yelp dataset is a subset of Yelp's businesses, reviews, and user data. The Amazon dataset is a subset of customers' ratings for Amazon books. Considering the ratings are integers ranging from 1 to 5, the ratings above 4 are regarded as positive. Following \cite{10.1145/3366423.3380187, 10.1145/3485447.3512075}, we leverage the routine strategy — 5-core setting to preprocess the dataset. \par
For each user, we randomly select 80\% of items to form the training set and 20\% of items to form the test set. 10\% of the training set is used for validation. The models are built on the training set and evaluated on the test set. 

\begin{table}
  \caption{The Statistics of Datasets}
  \label{tab:dataset}
  \Description{The Statistics of Datasets.}
  \begin{tabular}{c c c c c c}
    \toprule
    Dataset &User    &Item    &Train     &Test     &Sparsity\\
    \midrule
    Gowalla &29,858 &40,988 &822,358   &205,106 &99.9160\% \\
    Yelp    &77,277 &45,638 &1,684,846   &419,049 &99.9403\% \\
    Amazon &130,380 &128,939 &1,934,404   &481,246 &99.9856\% \\
  \bottomrule
\end{tabular}
\end{table}

\paragraph{Metrics}
When evaluating the models, we filter out positive items in the training set and utilize widely-used metrics Recall@$K$ and NDCG@$K$ to evaluate the recommendation performance. The detailed definitions are shown in Appendix \ref{Appendix_metrics}.

\begin{table*}[t]

\caption{Performance comparison on three datasets. The best results are in bold and the second best are underlined. The baselines are taken from \cite{10.1145/3485447.3512075}, as we completely follow their experiment settings. ``**'' denote the improvement is significant with t-test with $p<0.05$.}
\Description{Performance comparison on three datasets. Our methods outperform all baselines.}
\label{main_table}
\begin{tabular}{c|cc|cc|cc}
\hline
\multirow{2}{*}{Method} & \multicolumn{2}{c|}{Gowalla}  & \multicolumn{2}{c|}{Yelp}  & \multicolumn{2}{c}{Amazon} \\ 
\cline{2-7}
        & NDCG@50    & Recall@50       & NDCG@50    & Recall@50    & NDCG@50    & Recall@50    \\ \hline
BPR     & 0.1216         & 0.2048      & 0.0524   &  0.1083        &0.0499        & 0.1171     \\
AOBPR   & 0.1385         & 0.2417      & 0.0677   &  0.1346        & 0.0563       &  0.1303     \\
WARP   & 0.1248         &  0.2240      & 0.0636   &   0.1332       &  0.0542      & 0.1267     \\
IRGAN  &  0.1443        & 0.2242       & 0.0695   &  0.1367        & 0.0627       &  0.1395     \\
Kernel &  0.1399  & 0.2264    & 0.0658   &0.1315   &0.0700  &0.1495 \\
DNS  & 0.1412         & 0.1839       & 0.0693   &  0.1425        & 0.0615       &   0.1378     \\
PRIS(U) & 0.1334  &0.2217     &0.0639  &0.1273  & 0.0607         & 0.1377        \\
PRIS(P) &0.1385    & 0.2282    &0.0673  &0.1342      & 0.0697        & 0.1463   \\
AdaSIR(U)  &0.1489 &0.2500     &0.0732  &0.1523       & 0.0731         &0.1505  \\
AdaSIR(P)  &0.1519 & 0.2516      &0.0731   &0.1525  &0.0740         & 0.1534      \\
\hline
DNS($M$, $N$)  &\underline{0.1811**}   & \underline{0.2989**}       &\textbf{0.0899**}  &\textbf{0.1774**}         &\underline{0.1014**}        & \underline{0.1833**}     \\
Softmax-v($\rho$, $N$)  &\textbf{0.1837**}  & \textbf{0.2993**}      &\underline{0.0840**}   &\underline{0.1690**}  &\textbf{0.1046**}          &\textbf{0.1937**}        \\
\hline
\end{tabular}

\end{table*}

\subsection{Baselines}
To verify the effectiveness of DNS($M$, $N$) and Softmax-v($\rho$, $N$) methods, we compare our algorithms with the following baselines. 
\begin{itemize}[leftmargin=*]
    \item \textbf{BPR} \cite{10.5555/1795114.1795167} is a classical method for implicit feedback. It utilizes pairwise logit loss and randomly samples negative items.
    \item \textbf{AOBPR} \cite{10.1145/2556195.2556248} improves BPR through adaptively oversampling top-ranked negative items.
    \item \textbf{WARP} \cite{10.5555/2283696.2283856} uses the Weighted Approximate-Rank Pairwise loss function for implicit feedback.
    \item \textbf{IRGAN} \cite{10.1145/3077136.3080786} utilizes a minimax game to optimize the generative and discriminative network simultaneously. The negative items are drawn based on softmax distribution.
    \item \textbf{DNS} \cite{10.1145/2484028.2484126} is a dynamic negative sampler, which is a special case of DNS($M$, $N$) with $M=1$. 
    \item \textbf{Kernel} \cite{pmlr-v80-blanc18a} is an efficient sampling method that approximates the softmax distribution with non-negative quadratic kernel. 
    \item \textbf{PRIS} \cite{10.1145/3366423.3380187} utilizes importance sampling for training, where importance weights are based on softmax distribution. They adopt the uniform and popularity-based distribution to construct the sampling pool, denoted as PRIS(U) and PRIS(P), respectively.
    \item \textbf{AdaSIR} \cite{10.1145/3485447.3512075} is a two-stage method that maintains a fixed size contextualized sample pool with importance resampling. The importance weights are based on softmax distribution. They adopt the uniform and popularity-based distribution to construct the sampling pool, denoted as AdaSIR(U) and AdaSIR(P), respectively.
\end{itemize}

\subsection{Implementation Details}\label{implementaion_detail}
The algorithms are implemented based on PyTorch. We completely follow the experiments setting in \cite{10.1145/3366423.3380187, 10.1145/3485447.3512075}. We utilize Matrix Factorization (MF) as the recommender model for our model. We utilize Adam optimizer to optimize all parameters. The dimension of user and item embedding is set to 32. The batch size is fixed to 4096, and the learning rate is set to 0.001 by default. The number of training epochs is set to 200 for all methods. We utilize grid search to find the best with weight\_decay $\in$ \{0.1, 0.01, 0.001, 0.0001\}. The hyperparameter $M$ in DNS($M$, $N$) is tuned over \{1,2,3,4,5,10,20\} and the hyperparameter $\rho$ of Softmax-v($\rho$, $N$) is tuned over \{0.01, 0.1, 1, 10, 100\} for all datasets. Due to the efficiency limit, the sample pool size $N$ for each user is set to 200, 200, and 500 for Gowalla, Yelp, and Amazon. The maximum number of negative samples per positive pair $(c, i)$ is the sample pool size. The baseline results are directly taken from \cite{10.1145/3485447.3512075}, as we completely follow their experiment setting. Code is available at https://github.com/swt-user/WWW\_2023\_code. \par

\subsection{(RQ1) Performance Comparison}
Table \ref{main_table} shows the performance of DNS($M$, $N$), Softmax-v($\rho$, $N$), and baselines. From them, we have the following key findings:
\begin{itemize}[leftmargin=*]
    \item Compared to the uniform negative sampling method BPR, most HNS methods perform much better, especially DNS($M$, $N$) and Softmax-v($\rho$, $N$). This clearly verifies the effectiveness of HNS.
    \item Benefiting from the adjustable sampling hardness, DNS($M$, $N$) significantly outperform its original version on average 40\%. Meanwhile, the two methods also present a huge performance boost over other HNS methods. These findings demonstrate the extreme importance of our first guideline in Section \ref{section5}.
\end{itemize}

\begin{figure}[t]
  \centering
  \includegraphics[width=0.95\linewidth]{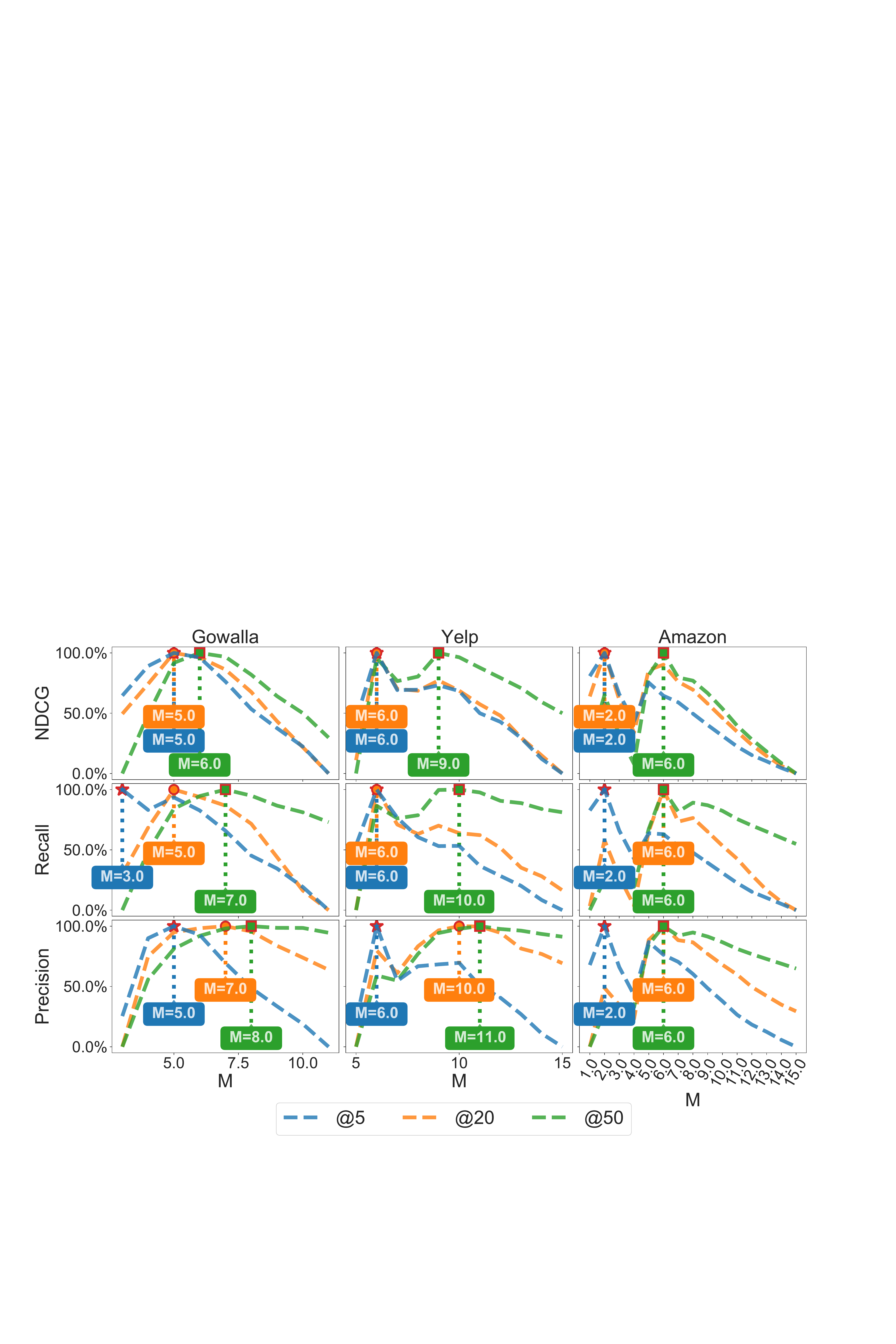}
  \caption{The effect of $M$ in DNS($M$, $N$), where $N$ is set to 200, 200, 500 for Gowalla, Yelp and Amazon respectively.}
  \Description{For all datasets and all metrics, the lower the $K$ in Top-$K$ metrics is, the smaller the $M$ in DNS($M$, $N$) when the curve achieves its maximum performance.}
  \label{fig:top_k}
\end{figure}

\begin{figure}[t]
  \centering
  \includegraphics[width=0.95\linewidth]{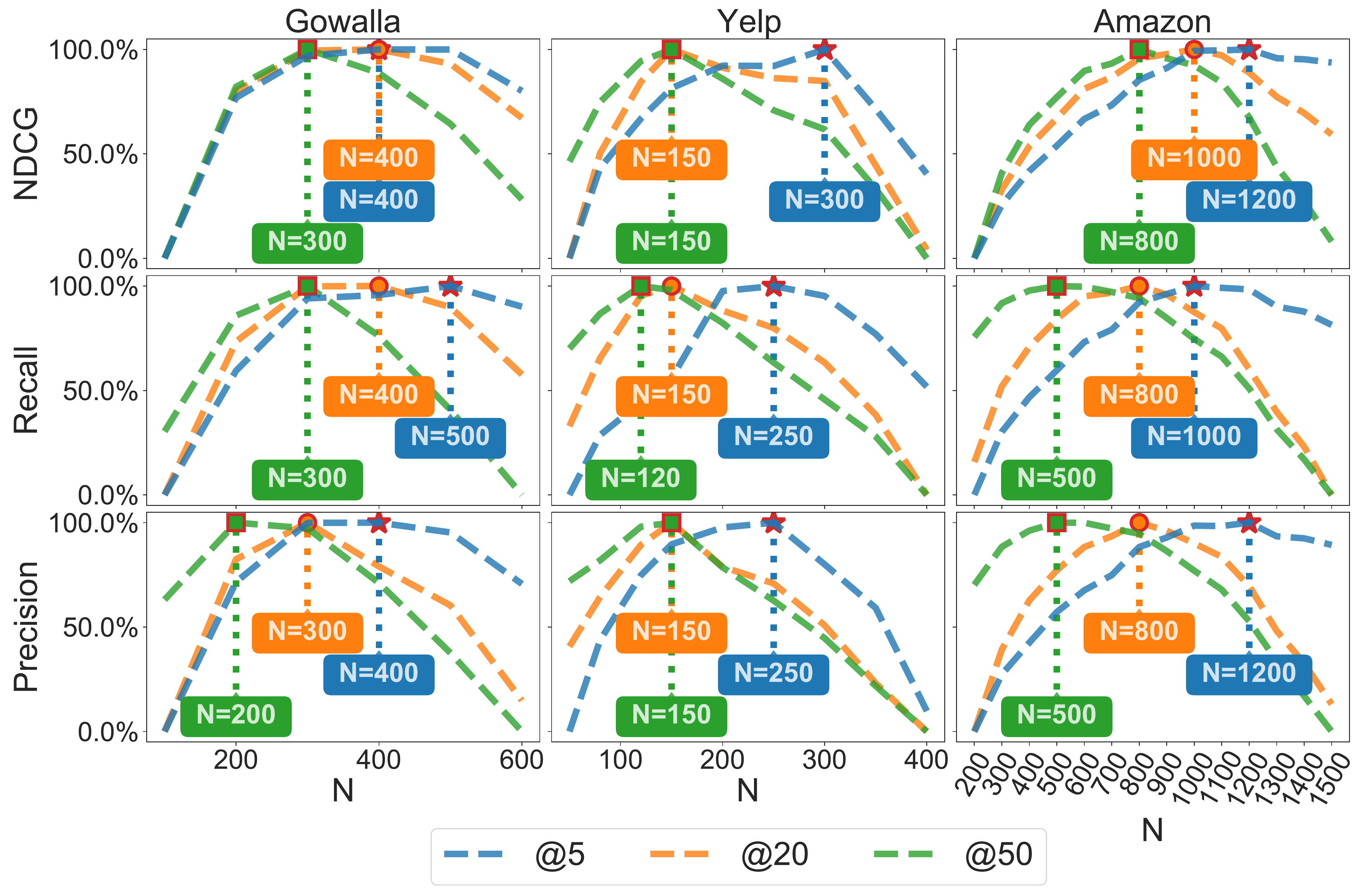}
  \caption{The effect of $N$ in DNS($M$, $N$), where $M$ is set to 5 for all three datasets.}
  \Description{For all datasets and all metrics, the lower the $K$ in Top-$K$ metrics is, the larger the $N$ in DNS($M$, $N$) when the curve achieves its maximum performance.}
  \label{fig:pool_size}
\end{figure}

\begin{figure}[t]
  \centering
  \includegraphics[width=0.95\linewidth]{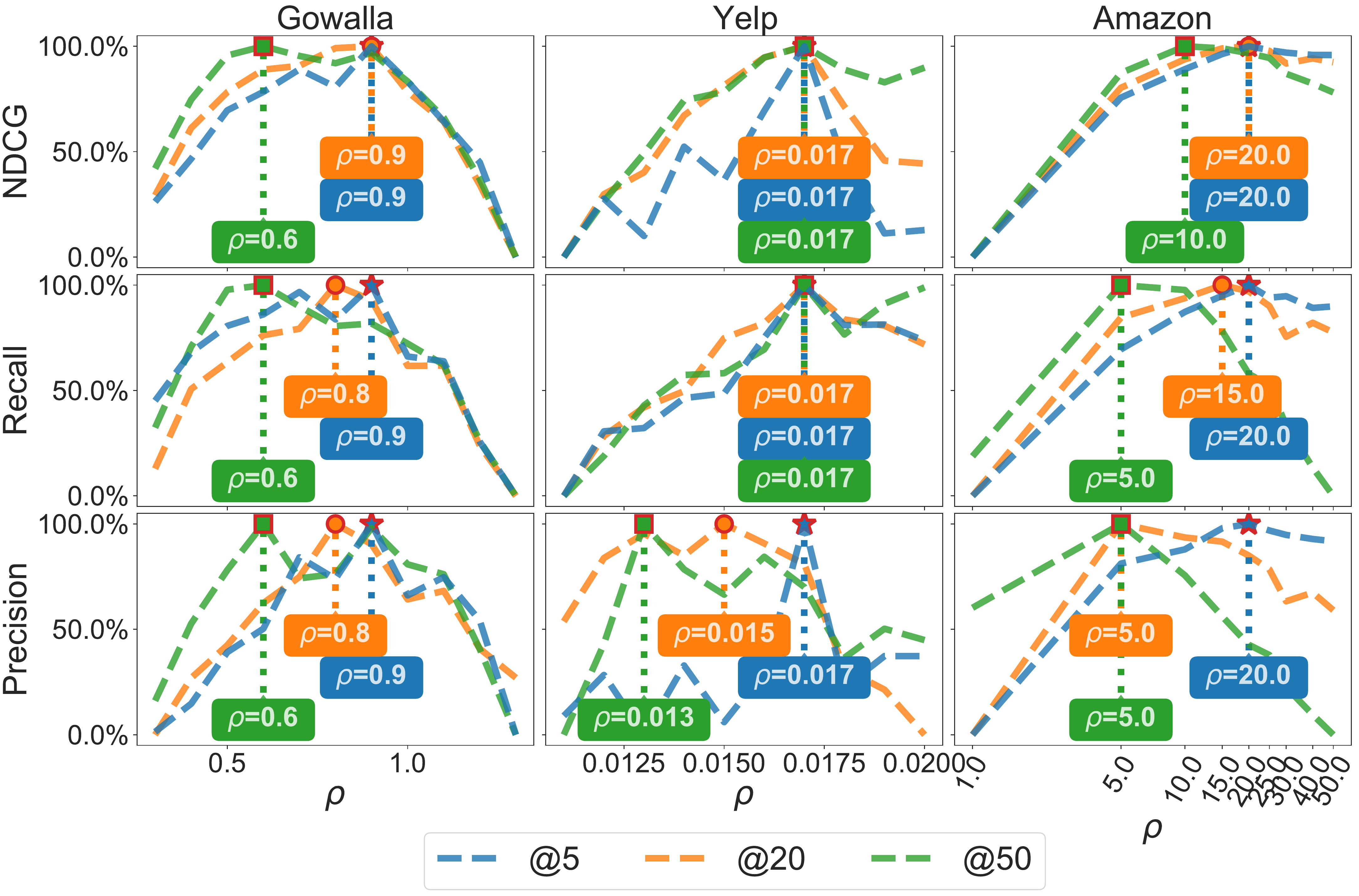}
  \caption{The effect of $\rho$ in Softmax-v($\rho$, $N$), where $N$ is set to 200, 200, 500 for Gowalla, Yelp and Amazon respectively.}
  \Description{For all datasets and all metrics, the lower the $K$ in Top-$K$ metrics is, the larger the $\rho$ in Softmax-v($\rho$, $N$) when the curve achieves its maximum performance.}
  \label{fig:softmax}
\end{figure}

\subsection{(RQ2) Performance with Different Sampling Distributions}
This subsection investigates how Top-$K$ metrics will change under different sampling distributions on real-world datasets. As the sampling distribution is affected by hyperparameters, see Section \ref{section5}, we investigate the performance under different hyperparameter settings. \par
We report the performance results on three public dataset under different $M$ in DNS(M,N), different N in DNS(M, N) and different $\rho$ in Softmax-v($\rho$,N) in Figure \ref{fig:top_k}, Figure \ref{fig:pool_size} and Figure \ref{fig:softmax} respectively. We only care about the relative magnitude of Top-$K$ metrics, so we report the relative value of Top-$K$ evaluation metrics for better visualization. We highlight the value of hyperparameters when each curve reaches its maximum result. For each result, we tune the learning rate $\in$ \{0.01, 0.001\} and weight\_decay $\in$ \{0.01, 0.001, 0.0001\} to find the best. \par
\begin{itemize}[leftmargin=*]
    \item From Figure \ref{fig:top_k}, we observe that for all datasets and all metrics, the lower the $K$ in Top-$K$ metrics is, the smaller the $M$ in DNS($M$, $N$) when the curve achieves its maximum performance.
    \item From Figure \ref{fig:pool_size}, we observe that for all datasets and all metrics, the lower the $K$ in Top-$K$ metrics is, the larger the $N$ in DNS($M$, $N$) when the curve achieves its maximum performance.
    \item From Figure \ref{fig:softmax}, we observe that for all datasets and all metrics, the lower the $K$ in Top-$K$ metrics is, the larger the $\rho$ in Softmax-v($\rho$, $N$) when the curve achieves its maximum performance.
\end{itemize}
In some cases, the peak of the Top-20 curve coincides with the peak of the Top-50 curve or Top-5 curve. This can be attributed to the relatively small difference of $K$. With a larger difference of $K$, for example, Top-50 and Top-5, their curve always matches our observation. We conduct further experiments to investigate the performance across a wide range of $K$ in Appendix \ref{appendix_further_K}. \par
Recall that we have observed how hyperparameters ($M$, $N$, $\rho$) affect sampling hardness in Figure \ref{Cumulative_Probability}. Combining these two observations, we can easily conclude that \textbf{the smaller the $K$ in Top-$K$ metrics, the harder the negative samples we should draw.} These clearly verify our second guideline.

\balance

\section{Related Work}\label{section7}

\subsection{Negative Sampling for Recommendation}
 Early work sample items based on predefined distributions, e.g., uniform distribution \cite{ijcai2019-309, 10.5555/1795114.1795167} and popularity-based distribution \cite{10.1145/3240323.3240377, 10.1145/3097983.3098202}. These static samplers are independent of model status and unchanged for different users. Thus, the performance is limited. Later on, adaptive samplers are proposed, such as DNS \cite{10.1145/2484028.2484126} and softmax-based sampling methods. Softmax-based sampling is widely used in adversarial learning (e.g. IRGAN \cite{10.1145/3077136.3080786} and ADVIR \cite{10.1145/3308558.3313416}) and importance sampling (e.g. PRIS \cite{10.1145/3366423.3380187} and AdaSIR \cite{10.1145/3485447.3512075}). They assign high sampling probability to top-ranked negative items, accounting for model status. There are also some fine-grained negative sampling methods \cite{DBLP:conf/www/ZhuZHD22, DBLP:conf/www/Wan00WGT22, DBLP:conf/www/MaoWWL21, DBLP:conf/www/WangX000C20}. Empirical experiments verify the effectiveness and efficiency of HNS. The efficiency problem has been studied in AOBPR \cite{10.1145/2556195.2556248}. They argue that HNS samples more informative high-scored items, which can contribute more to the gradients and accelerate the convergence. Nevertheless, the reasons for the effectiveness of HNS are not revealed yet. To the best of our knowledge, only DNS \cite{10.1145/2484028.2484126} provides clues of the connection between HNS and Top-$K$ metrics. But unfortunately, they fail to give a theoretical foundation and deep analyses.

\subsection{Partial AUC Maximization}
Early work does not directly optimize the surrogate objective of Partial AUC, but instead, some other related objectives, e.g., p-norm push \cite{10.5555/1577069.1755861}, infinite-push \cite{ 10.5555/2968826.2968994}, and asymmetric SVM objective \cite{10.1145/1401890.1401980}. Nevertheless, these algorithms are not scalable and applicable to deep learning. More recently, \cite{conf/icml/0001XBHCH21} considers two-way partial AUC maximization and simplifies the optimizing problem for large scale optimization. \cite{pmlr-v162-zhu22g} proposes new formulations of Partial AUC surrogate objectives using distributionally robust optimization (DRO). This work motivates our proof of the connection between OPAUC and HNS. A more comprehensive study of AUC can refer to \cite{10.1145/3554729}.

\section{Conclusion}\label{section8}

In this paper, we reveal the theories behind HNS for recommendation. We prove that the BPR loss equipped with HNS strategies optimizes OPAUC. Meanwhile, we conduct theoretical analysis and simulation experiments to show the strong connection between OPAUC and Top-$K$ evaluation metrics. On these bases, the effectiveness of HNS can be clearly explained. To take a step further, we propose two insightful guidelines for effective usage of HNS.
In conclusion, the proposed theoretical understanding of HNS can not only explain effectiveness but also provide insightful guidelines for future study.\par
In the future, we will investigate the connection between Two-way Partial AUC and recommendation algorithms, which may bring more insights into recommendation systems. The effect of false negative items will also be an exciting research direction.

\begin{acks}
This work is supported by the National Key Research and Development Program of China (2022YFB3104701), the Starry Night Science Fund of Zhejiang University Shanghai Institute for Advanced Study (SN-ZJU-SIAS-001), the National Natural Science Foundation of China (61972372, 62121002, 62102382), and the CCCD Key Lab of Ministry of Culture and Tourism.
\end{acks}
\bibliographystyle{ACM-Reference-Format}
\bibliography{reference}


\begin{thebibliography}{42}


\ifx \showCODEN    \undefined \def \showCODEN     #1{\unskip}     \fi
\ifx \showDOI      \undefined \def \showDOI       #1{#1}\fi
\ifx \showISBNx    \undefined \def \showISBNx     #1{\unskip}     \fi
\ifx \showISBNxiii \undefined \def \showISBNxiii  #1{\unskip}     \fi
\ifx \showISSN     \undefined \def \showISSN      #1{\unskip}     \fi
\ifx \showLCCN     \undefined \def \showLCCN      #1{\unskip}     \fi
\ifx \shownote     \undefined \def \shownote      #1{#1}          \fi
\ifx \showarticletitle \undefined \def \showarticletitle #1{#1}   \fi
\ifx \showURL      \undefined \def \showURL       {\relax}        \fi
\providecommand\bibfield[2]{#2}
\providecommand\bibinfo[2]{#2}
\providecommand\natexlab[1]{#1}
\providecommand\showeprint[2][]{arXiv:#2}

\bibitem[Bayer et~al\mbox{.}(2017)]%
        {DBLP:conf/www/BayerHKR17}
\bibfield{author}{\bibinfo{person}{Immanuel Bayer}, \bibinfo{person}{Xiangnan
  He}, \bibinfo{person}{Bhargav Kanagal}, {and} \bibinfo{person}{Steffen
  Rendle}.} \bibinfo{year}{2017}\natexlab{}.
\newblock \showarticletitle{A Generic Coordinate Descent Framework for Learning
  from Implicit Feedback}. In \bibinfo{booktitle}{\emph{{WWW}}}.
  \bibinfo{publisher}{{ACM}}, \bibinfo{pages}{1341--1350}.
\newblock


\bibitem[Blanc and Rendle(2018)]%
        {pmlr-v80-blanc18a}
\bibfield{author}{\bibinfo{person}{Guy Blanc} {and} \bibinfo{person}{Steffen
  Rendle}.} \bibinfo{year}{2018}\natexlab{}.
\newblock \showarticletitle{Adaptive Sampled Softmax with Kernel Based
  Sampling}. In \bibinfo{booktitle}{\emph{ICML}}. \bibinfo{pages}{590--599}.
\newblock


\bibitem[Caselles{-}Dupr{\'{e}} et~al\mbox{.}(2018)]%
        {10.1145/3240323.3240377}
\bibfield{author}{\bibinfo{person}{Hugo Caselles{-}Dupr{\'{e}}},
  \bibinfo{person}{Florian Lesaint}, {and} \bibinfo{person}{Jimena
  Royo{-}Letelier}.} \bibinfo{year}{2018}\natexlab{}.
\newblock \showarticletitle{Word2vec applied to recommendation: hyperparameters
  matter}. In \bibinfo{booktitle}{\emph{RecSys}}. \bibinfo{pages}{352--356}.
\newblock


\bibitem[Chen et~al\mbox{.}(2022b)]%
        {10.1145/3522672}
\bibfield{author}{\bibinfo{person}{Chong Chen}, \bibinfo{person}{Weizhi Ma},
  \bibinfo{person}{Min Zhang}, \bibinfo{person}{Chenyang Wang},
  \bibinfo{person}{Yiqun Liu}, {and} \bibinfo{person}{Shaoping Ma}.}
  \bibinfo{year}{2022}\natexlab{b}.
\newblock \showarticletitle{Revisiting Negative Sampling VS. Non-Sampling in
  Implicit Recommendation}.
\newblock \bibinfo{journal}{\emph{ACM Trans. Inf. Syst.}}
  (\bibinfo{year}{2022}).
\newblock


\bibitem[Chen et~al\mbox{.}(2020b)]%
        {chen2020efficient}
\bibfield{author}{\bibinfo{person}{Chong Chen}, \bibinfo{person}{Min Zhang},
  \bibinfo{person}{Weizhi Ma}, \bibinfo{person}{Yongfeng Zhang},
  \bibinfo{person}{Yiqun Liu}, {and} \bibinfo{person}{Shaoping Ma}.}
  \bibinfo{year}{2020}\natexlab{b}.
\newblock \showarticletitle{Efﬁcient Heterogeneous Collaborative Filtering
  without Negative Sampling for Recommendation}. In
  \bibinfo{booktitle}{\emph{AAAI}}.
\newblock


\bibitem[Chen et~al\mbox{.}(2020c)]%
        {10.1145/3373807}
\bibfield{author}{\bibinfo{person}{Chong Chen}, \bibinfo{person}{Min Zhang},
  \bibinfo{person}{Yongfeng Zhang}, \bibinfo{person}{Yiqun Liu}, {and}
  \bibinfo{person}{Shaoping Ma}.} \bibinfo{year}{2020}\natexlab{c}.
\newblock \showarticletitle{Efficient Neural Matrix Factorization without
  Sampling for Recommendation}.
\newblock \bibinfo{journal}{\emph{ACM Trans. Inf. Syst.}}  \bibinfo{volume}{38}
  (\bibinfo{year}{2020}).
\newblock


\bibitem[Chen et~al\mbox{.}(2020a)]%
        {DBLP:journals/corr/abs-2010-03240}
\bibfield{author}{\bibinfo{person}{Jiawei Chen}, \bibinfo{person}{Hande Dong},
  \bibinfo{person}{Xiang Wang}, \bibinfo{person}{Fuli Feng},
  \bibinfo{person}{Meng Wang}, {and} \bibinfo{person}{Xiangnan He}.}
  \bibinfo{year}{2020}\natexlab{a}.
\newblock \showarticletitle{Bias and Debias in Recommender System: {A} Survey
  and Future Directions}.
\newblock \bibinfo{journal}{\emph{CoRR}}  \bibinfo{volume}{abs/2010.03240}
  (\bibinfo{year}{2020}).
\newblock


\bibitem[Chen et~al\mbox{.}(2021)]%
        {10.1145/3450289}
\bibfield{author}{\bibinfo{person}{Jiawei Chen}, \bibinfo{person}{Chengquan
  Jiang}, \bibinfo{person}{Can Wang}, \bibinfo{person}{Sheng Zhou},
  \bibinfo{person}{Yan Feng}, \bibinfo{person}{Chun Chen},
  \bibinfo{person}{Martin Ester}, {and} \bibinfo{person}{Xiangnan He}.}
  \bibinfo{year}{2021}\natexlab{}.
\newblock \showarticletitle{CoSam: An Efficient Collaborative Adaptive Sampler
  for Recommendation}.
\newblock \bibinfo{journal}{\emph{ACM Trans. Inf. Syst.}}  \bibinfo{volume}{39}
  (\bibinfo{year}{2021}).
\newblock


\bibitem[Chen et~al\mbox{.}(2022a)]%
        {10.1145/3485447.3512075}
\bibfield{author}{\bibinfo{person}{Jin Chen}, \bibinfo{person}{Defu Lian},
  \bibinfo{person}{Binbin Jin}, \bibinfo{person}{Kai Zheng}, {and}
  \bibinfo{person}{Enhong Chen}.} \bibinfo{year}{2022}\natexlab{a}.
\newblock \showarticletitle{Learning Recommenders for Implicit Feedback with
  Importance Resampling}. In \bibinfo{booktitle}{\emph{WWW}}.
  \bibinfo{pages}{1997--2005}.
\newblock


\bibitem[Chen et~al\mbox{.}(2017)]%
        {10.1145/3097983.3098202}
\bibfield{author}{\bibinfo{person}{Ting Chen}, \bibinfo{person}{Yizhou Sun},
  \bibinfo{person}{Yue Shi}, {and} \bibinfo{person}{Liangjie Hong}.}
  \bibinfo{year}{2017}\natexlab{}.
\newblock \showarticletitle{On Sampling Strategies for Neural Network-based
  Collaborative Filtering}. In \bibinfo{booktitle}{\emph{SIGKDD}}.
\newblock


\bibitem[Ding et~al\mbox{.}(2019)]%
        {ijcai2019-309}
\bibfield{author}{\bibinfo{person}{Jingtao Ding}, \bibinfo{person}{Yuhan Quan},
  \bibinfo{person}{Xiangnan He}, \bibinfo{person}{Yong Li}, {and}
  \bibinfo{person}{Depeng Jin}.} \bibinfo{year}{2019}\natexlab{}.
\newblock \showarticletitle{Reinforced Negative Sampling for Recommendation
  with Exposure Data}. In \bibinfo{booktitle}{\emph{IJCAI}}.
  \bibinfo{pages}{2230--2236}.
\newblock


\bibitem[Ding et~al\mbox{.}(2020)]%
        {NEURIPS2020_0c7119e3}
\bibfield{author}{\bibinfo{person}{Jingtao Ding}, \bibinfo{person}{Yuhan Quan},
  \bibinfo{person}{Quanming Yao}, \bibinfo{person}{Yong Li}, {and}
  \bibinfo{person}{Depeng Jin}.} \bibinfo{year}{2020}\natexlab{}.
\newblock \showarticletitle{Simplify and Robustify Negative Sampling for
  Implicit Collaborative Filtering}. In \bibinfo{booktitle}{\emph{NIPS}}.
  \bibinfo{pages}{1094--1105}.
\newblock


\bibitem[Dodd and Pepe(2003)]%
        {10.2307/3695437}
\bibfield{author}{\bibinfo{person}{Lori~E Dodd} {and}
  \bibinfo{person}{Margaret~S Pepe}.} \bibinfo{year}{2003}\natexlab{}.
\newblock \showarticletitle{Partial AUC estimation and regression}.
\newblock \bibinfo{journal}{\emph{Biometrics}} \bibinfo{volume}{59},
  \bibinfo{number}{3} (\bibinfo{year}{2003}), \bibinfo{pages}{614--623}.
\newblock


\bibitem[Duchi and Namkoong(2018)]%
        {1810.08750}
\bibfield{author}{\bibinfo{person}{John~C. Duchi} {and}
  \bibinfo{person}{Hongseok Namkoong}.} \bibinfo{year}{2018}\natexlab{}.
\newblock \showarticletitle{Learning Models with Uniform Performance via
  Distributionally Robust Optimization}.
\newblock \bibinfo{journal}{\emph{CoRR}}  \bibinfo{volume}{abs/1810.08750}
  (\bibinfo{year}{2018}).
\newblock


\bibitem[Faury et~al\mbox{.}(2020)]%
        {Faury_Tanielian_Dohmatob_Smirnova_Vasile_2020}
\bibfield{author}{\bibinfo{person}{Louis Faury}, \bibinfo{person}{Ugo
  Tanielian}, \bibinfo{person}{Elvis Dohmatob}, \bibinfo{person}{Elena
  Smirnova}, {and} \bibinfo{person}{Flavian Vasile}.}
  \bibinfo{year}{2020}\natexlab{}.
\newblock \showarticletitle{Distributionally Robust Counterfactual Risk
  Minimization}. In \bibinfo{booktitle}{\emph{AAAI}}.
  \bibinfo{pages}{3850--3857}.
\newblock


\bibitem[Gao and Zhou(2015)]%
        {10.5555/2832249.2832379}
\bibfield{author}{\bibinfo{person}{Wei Gao} {and} \bibinfo{person}{Zhi{-}Hua
  Zhou}.} \bibinfo{year}{2015}\natexlab{}.
\newblock \showarticletitle{On the Consistency of {AUC} Pairwise Optimization}.
  In \bibinfo{booktitle}{\emph{IJCAI}}. \bibinfo{pages}{939--945}.
\newblock


\bibitem[He et~al\mbox{.}(2016)]%
        {DBLP:conf/sigir/HeZKC16}
\bibfield{author}{\bibinfo{person}{Xiangnan He}, \bibinfo{person}{Hanwang
  Zhang}, \bibinfo{person}{Min{-}Yen Kan}, {and} \bibinfo{person}{Tat{-}Seng
  Chua}.} \bibinfo{year}{2016}\natexlab{}.
\newblock \showarticletitle{Fast Matrix Factorization for Online Recommendation
  with Implicit Feedback}. In \bibinfo{booktitle}{\emph{{SIGIR}}}.
  \bibinfo{publisher}{{ACM}}, \bibinfo{pages}{549--558}.
\newblock


\bibitem[Hu and Hong(2013)]%
        {Hu2012KullbackLeiblerDC}
\bibfield{author}{\bibinfo{person}{Zhaolin Hu} {and} \bibinfo{person}{L~Jeff
  Hong}.} \bibinfo{year}{2013}\natexlab{}.
\newblock \showarticletitle{Kullback-Leibler divergence constrained
  distributionally robust optimization}.
\newblock \bibinfo{journal}{\emph{Available at Optimization Online}}
  (\bibinfo{year}{2013}), \bibinfo{pages}{1695--1724}.
\newblock


\bibitem[Levy et~al\mbox{.}(2020)]%
        {NEURIPS2020_64986d86}
\bibfield{author}{\bibinfo{person}{Daniel Levy}, \bibinfo{person}{Yair Carmon},
  \bibinfo{person}{John~C Duchi}, {and} \bibinfo{person}{Aaron Sidford}.}
  \bibinfo{year}{2020}\natexlab{}.
\newblock \showarticletitle{Large-Scale Methods for Distributionally Robust
  Optimization}. In \bibinfo{booktitle}{\emph{NIPS}},
  Vol.~\bibinfo{volume}{33}. \bibinfo{pages}{8847--8860}.
\newblock


\bibitem[Li et~al\mbox{.}(2014)]%
        {10.5555/2968826.2968994}
\bibfield{author}{\bibinfo{person}{Nan Li}, \bibinfo{person}{Rong Jin}, {and}
  \bibinfo{person}{Zhi{-}Hua Zhou}.} \bibinfo{year}{2014}\natexlab{}.
\newblock \showarticletitle{Top Rank Optimization in Linear Time}. In
  \bibinfo{booktitle}{\emph{{NIPS}}}. \bibinfo{pages}{1502--1510}.
\newblock


\bibitem[Lian et~al\mbox{.}(2020)]%
        {10.1145/3366423.3380187}
\bibfield{author}{\bibinfo{person}{Defu Lian}, \bibinfo{person}{Qi Liu}, {and}
  \bibinfo{person}{Enhong Chen}.} \bibinfo{year}{2020}\natexlab{}.
\newblock \showarticletitle{Personalized Ranking with Importance Sampling}. In
  \bibinfo{booktitle}{\emph{WWW}}. \bibinfo{pages}{1093--1103}.
\newblock


\bibitem[Lin et~al\mbox{.}(2022)]%
        {2155-3289_2022_1_159}
\bibfield{author}{\bibinfo{person}{Fengming Lin}, \bibinfo{person}{Xiaolei
  Fang}, {and} \bibinfo{person}{Zheming Gao}.} \bibinfo{year}{2022}\natexlab{}.
\newblock \showarticletitle{Distributionally Robust Optimization: A review on
  theory and applications}.
\newblock \bibinfo{journal}{\emph{Numerical Algebra, Control and Optimization}}
   \bibinfo{volume}{12} (\bibinfo{year}{2022}), \bibinfo{pages}{159--212}.
\newblock


\bibitem[Mao et~al\mbox{.}(2021)]%
        {DBLP:conf/www/MaoWWL21}
\bibfield{author}{\bibinfo{person}{Xin Mao}, \bibinfo{person}{Wenting Wang},
  \bibinfo{person}{Yuanbin Wu}, {and} \bibinfo{person}{Man Lan}.}
  \bibinfo{year}{2021}\natexlab{}.
\newblock \showarticletitle{Boosting the Speed of Entity Alignment 10
  {\texttimes}: Dual Attention Matching Network with Normalized Hard Sample
  Mining}. In \bibinfo{booktitle}{\emph{{WWW}}}. \bibinfo{pages}{821--832}.
\newblock


\bibitem[McClish(1989)]%
        {article_Analy}
\bibfield{author}{\bibinfo{person}{Donna McClish}.}
  \bibinfo{year}{1989}\natexlab{}.
\newblock \showarticletitle{Analyzing a portion of the ROC Curve}.
\newblock \bibinfo{journal}{\emph{Medical decision making : an international
  journal of the Society for Medical Decision Making}}  \bibinfo{volume}{9}
  (\bibinfo{year}{1989}), \bibinfo{pages}{190--5}.
\newblock


\bibitem[Namdar et~al\mbox{.}(2021)]%
        {DBLP:journals/frai/NamdarHK21}
\bibfield{author}{\bibinfo{person}{Khashayar Namdar},
  \bibinfo{person}{Masoom~A. Haider}, {and} \bibinfo{person}{Farzad Khalvati}.}
  \bibinfo{year}{2021}\natexlab{}.
\newblock \showarticletitle{A Modified {AUC} for Training Convolutional Neural
  Networks: Taking Confidence Into Account}.
\newblock \bibinfo{journal}{\emph{Frontiers Artif. Intell.}}
  \bibinfo{volume}{4} (\bibinfo{year}{2021}), \bibinfo{pages}{582928}.
\newblock


\bibitem[Park and Chang(2019)]%
        {10.1145/3308558.3313416}
\bibfield{author}{\bibinfo{person}{Dae~Hoon Park} {and} \bibinfo{person}{Yi
  Chang}.} \bibinfo{year}{2019}\natexlab{}.
\newblock \showarticletitle{Adversarial Sampling and Training for
  Semi-Supervised Information Retrieval}. In \bibinfo{booktitle}{\emph{WWW}}.
  \bibinfo{pages}{1443–1453}.
\newblock


\bibitem[Rahimian and Mehrotra(2019)]%
        {1908.05659}
\bibfield{author}{\bibinfo{person}{Hamed Rahimian} {and}
  \bibinfo{person}{Sanjay Mehrotra}.} \bibinfo{year}{2019}\natexlab{}.
\newblock \showarticletitle{Distributionally Robust Optimization: {A} Review}.
\newblock \bibinfo{journal}{\emph{CoRR}}  \bibinfo{volume}{abs/1908.05659}
  (\bibinfo{year}{2019}).
\newblock


\bibitem[Rendle and Freudenthaler(2014)]%
        {10.1145/2556195.2556248}
\bibfield{author}{\bibinfo{person}{Steffen Rendle} {and}
  \bibinfo{person}{Christoph Freudenthaler}.} \bibinfo{year}{2014}\natexlab{}.
\newblock \showarticletitle{Improving Pairwise Learning for Item Recommendation
  from Implicit Feedback}. In \bibinfo{booktitle}{\emph{WSDM}}.
  \bibinfo{pages}{273–282}.
\newblock


\bibitem[Rendle et~al\mbox{.}(2009)]%
        {10.5555/1795114.1795167}
\bibfield{author}{\bibinfo{person}{Steffen Rendle}, \bibinfo{person}{Christoph
  Freudenthaler}, \bibinfo{person}{Zeno Gantner}, {and} \bibinfo{person}{Lars
  Schmidt{-}Thieme}.} \bibinfo{year}{2009}\natexlab{}.
\newblock \showarticletitle{{BPR:} Bayesian Personalized Ranking from Implicit
  Feedback}. In \bibinfo{booktitle}{\emph{UAI}}. \bibinfo{pages}{452--461}.
\newblock


\bibitem[Rockafellar(2017)]%
        {rockafellar2017risk}
\bibfield{author}{\bibinfo{person}{R~Tyrrell Rockafellar}.}
  \bibinfo{year}{2017}\natexlab{}.
\newblock \showarticletitle{Risk and utility in the duality framework of convex
  analysis}. In \bibinfo{booktitle}{\emph{Jonathan M. Borwein Commemorative
  Conference}}. \bibinfo{pages}{21--42}.
\newblock


\bibitem[Rudin(2009)]%
        {10.5555/1577069.1755861}
\bibfield{author}{\bibinfo{person}{Cynthia Rudin}.}
  \bibinfo{year}{2009}\natexlab{}.
\newblock \showarticletitle{The P-Norm Push: A Simple Convex Ranking Algorithm
  That Concentrates at the Top of the List}.
\newblock \bibinfo{journal}{\emph{J. Mach. Learn. Res.}}  \bibinfo{volume}{10}
  (\bibinfo{year}{2009}), \bibinfo{pages}{2233–2271}.
\newblock


\bibitem[Wan et~al\mbox{.}(2022)]%
        {DBLP:conf/www/Wan00WGT22}
\bibfield{author}{\bibinfo{person}{Qi Wan}, \bibinfo{person}{Xiangnan He},
  \bibinfo{person}{Xiang Wang}, \bibinfo{person}{Jiancan Wu},
  \bibinfo{person}{Wei Guo}, {and} \bibinfo{person}{Ruiming Tang}.}
  \bibinfo{year}{2022}\natexlab{}.
\newblock \showarticletitle{Cross Pairwise Ranking for Unbiased Item
  Recommendation}. In \bibinfo{booktitle}{\emph{{WWW}}}.
  \bibinfo{pages}{2370--2378}.
\newblock


\bibitem[Wang et~al\mbox{.}(2017)]%
        {10.1145/3077136.3080786}
\bibfield{author}{\bibinfo{person}{Jun Wang}, \bibinfo{person}{Lantao Yu},
  \bibinfo{person}{Weinan Zhang}, \bibinfo{person}{Yu Gong},
  \bibinfo{person}{Yinghui Xu}, \bibinfo{person}{Benyou Wang},
  \bibinfo{person}{Peng Zhang}, {and} \bibinfo{person}{Dell Zhang}.}
  \bibinfo{year}{2017}\natexlab{}.
\newblock \showarticletitle{IRGAN: A Minimax Game for Unifying Generative and
  Discriminative Information Retrieval Models}. In
  \bibinfo{booktitle}{\emph{SIGIR}}. \bibinfo{pages}{515–524}.
\newblock


\bibitem[Wang et~al\mbox{.}(2020)]%
        {DBLP:conf/www/WangX000C20}
\bibfield{author}{\bibinfo{person}{Xiang Wang}, \bibinfo{person}{Yaokun Xu},
  \bibinfo{person}{Xiangnan He}, \bibinfo{person}{Yixin Cao},
  \bibinfo{person}{Meng Wang}, {and} \bibinfo{person}{Tat{-}Seng Chua}.}
  \bibinfo{year}{2020}\natexlab{}.
\newblock \showarticletitle{Reinforced Negative Sampling over Knowledge Graph
  for Recommendation}. In \bibinfo{booktitle}{\emph{{WWW}}}.
  \bibinfo{pages}{99--109}.
\newblock


\bibitem[Weston et~al\mbox{.}(2011)]%
        {10.5555/2283696.2283856}
\bibfield{author}{\bibinfo{person}{Jason Weston}, \bibinfo{person}{Samy
  Bengio}, {and} \bibinfo{person}{Nicolas Usunier}.}
  \bibinfo{year}{2011}\natexlab{}.
\newblock \showarticletitle{WSABIE: Scaling up to Large Vocabulary Image
  Annotation}. In \bibinfo{booktitle}{\emph{IJCAI}}.
  \bibinfo{pages}{2764–2770}.
\newblock


\bibitem[Wu et~al\mbox{.}(2008)]%
        {10.1145/1401890.1401980}
\bibfield{author}{\bibinfo{person}{Shan{-}Hung Wu}, \bibinfo{person}{Keng{-}Pei
  Lin}, \bibinfo{person}{Chung{-}Min Chen}, {and} \bibinfo{person}{Ming{-}Syan
  Chen}.} \bibinfo{year}{2008}\natexlab{}.
\newblock \showarticletitle{Asymmetric support vector machines: low
  false-positive learning under the user tolerance}. In
  \bibinfo{booktitle}{\emph{{KDD}}}. \bibinfo{publisher}{{ACM}},
  \bibinfo{pages}{749--757}.
\newblock


\bibitem[Yang and Ying(2022)]%
        {10.1145/3554729}
\bibfield{author}{\bibinfo{person}{Tianbao Yang} {and} \bibinfo{person}{Yiming
  Ying}.} \bibinfo{year}{2022}\natexlab{}.
\newblock \showarticletitle{AUC Maximization in the Era of Big Data and AI: A
  Survey}.
\newblock \bibinfo{journal}{\emph{ACM Comput. Surv.}} (\bibinfo{date}{jul}
  \bibinfo{year}{2022}).
\newblock


\bibitem[Yang et~al\mbox{.}(2021)]%
        {conf/icml/0001XBHCH21}
\bibfield{author}{\bibinfo{person}{Zhiyong Yang}, \bibinfo{person}{Qianqian
  Xu}, \bibinfo{person}{Shilong Bao}, \bibinfo{person}{Yuan He},
  \bibinfo{person}{Xiaochun Cao}, {and} \bibinfo{person}{Qingming Huang}.}
  \bibinfo{year}{2021}\natexlab{}.
\newblock \showarticletitle{When All We Need is a Piece of the Pie: A Generic
  Framework for Optimizing Two-way Partial AUC}. In
  \bibinfo{booktitle}{\emph{ICML}}. \bibinfo{pages}{11820--11829}.
\newblock


\bibitem[Zhai et~al\mbox{.}(2021)]%
        {pmlr-v139-zhai21a}
\bibfield{author}{\bibinfo{person}{Runtian Zhai}, \bibinfo{person}{Chen Dan},
  \bibinfo{person}{J.~Zico Kolter}, {and} \bibinfo{person}{Pradeep Ravikumar}.}
  \bibinfo{year}{2021}\natexlab{}.
\newblock \showarticletitle{{DORO:} Distributional and Outlier Robust
  Optimization}. In \bibinfo{booktitle}{\emph{{ICML}}}.
  \bibinfo{pages}{12345--12355}.
\newblock


\bibitem[Zhang et~al\mbox{.}(2013)]%
        {10.1145/2484028.2484126}
\bibfield{author}{\bibinfo{person}{Weinan Zhang}, \bibinfo{person}{Tianqi
  Chen}, \bibinfo{person}{Jun Wang}, {and} \bibinfo{person}{Yong Yu}.}
  \bibinfo{year}{2013}\natexlab{}.
\newblock \showarticletitle{Optimizing top-n collaborative filtering via
  dynamic negative item sampling}. In \bibinfo{booktitle}{\emph{SIGIR}}.
  \bibinfo{pages}{785--788}.
\newblock


\bibitem[Zhu et~al\mbox{.}(2022a)]%
        {pmlr-v162-zhu22g}
\bibfield{author}{\bibinfo{person}{Dixian Zhu}, \bibinfo{person}{Gang Li},
  \bibinfo{person}{Bokun Wang}, \bibinfo{person}{Xiaodong Wu}, {and}
  \bibinfo{person}{Tianbao Yang}.} \bibinfo{year}{2022}\natexlab{a}.
\newblock \showarticletitle{When {AUC} meets {DRO:} Optimizing Partial {AUC}
  for Deep Learning with Non-Convex Convergence Guarantee}. In
  \bibinfo{booktitle}{\emph{{ICML}}}. \bibinfo{pages}{27548--27573}.
\newblock


\bibitem[Zhu et~al\mbox{.}(2022b)]%
        {DBLP:conf/www/ZhuZHD22}
\bibfield{author}{\bibinfo{person}{Qiannan Zhu}, \bibinfo{person}{Haobo Zhang},
  \bibinfo{person}{Qing He}, {and} \bibinfo{person}{Zhicheng Dou}.}
  \bibinfo{year}{2022}\natexlab{b}.
\newblock \showarticletitle{A Gain-Tuning Dynamic Negative Sampler for
  Recommendation}. In \bibinfo{booktitle}{\emph{{WWW}}}.
  \bibinfo{pages}{277--285}.
\newblock


\end{thebibliography}


\newpage
\appendix
\section{Proof of Theorem \ref{Theorem_Softmax_OPAUC}} \label{proof_of_softmax}
\begin{proof}
As shown in Lemma \ref{lemma_CVAR_OPAUC}, the DRO-based objective (Eq. \eqref{DRO_object}) is equivalent to OPAUC($\beta$) (Eq. \eqref{object_OPAUC}). By replacing CVaR divergence with KL divergence $D_\phi=D_{KL}(Q||P_0) = \int \log(\frac{\mathrm{d}Q}{\mathrm{d}P}) \mathrm{d}Q$, then the DRO-based objective (Eq. \eqref{DRO_object})) reduces to
\begin{equation}
    \min_{\theta} \min_{\mathbf{\lambda}\ge 0} \frac{1}{|\mathcal{C}|} \sum_{c\in\mathcal{C}} \frac{1}{n_+} \sum_{i\in\mathcal{I}_c^+} \{\lambda_i \cdot \log E_{j\sim P_0} \left[\exp\left( \frac{L(c,i,j)}{\lambda_i} \right)\right]  +\lambda_i\cdot\rho \}.
    \label{softmax_proof}
\end{equation}
The detailed derivation can be found in \cite{Hu2012KullbackLeiblerDC}. By setting $\beta=\exp(-\rho)$, we get a surrogate objective of $OPAUC(\beta)$. Next, we will show that it is equivalent to the softmax-based sampling problem. \par
Differentiate the objective respect to $\lambda_i$ and set to 0, and then we find that the optimal $\lambda_i$ is the solution to the fixed-point equation:
\begin{equation}
    \lambda_i = E_{j\sim P_0} \left[\frac{e^{\frac{L(c,i,j)}{\lambda_i}}\cdot L(c,i,j)}{E_{j\sim P_0}[e^{\frac{L(c,i,j)}{\lambda_i}}]} \right] \cdot \frac{1}{\rho + \log E_{j\sim P_0}[e^{\frac{L(c,i,j)}{\lambda_i}}]}.
    \label{softmax_proof_2}
\end{equation}
Replace the above value for $\lambda_i$ in Eq. \eqref{softmax_proof}, and then we derive the following result: 
\begin{equation}
    \min_{\theta} \frac{1}{|\mathcal{C}|} \sum_{c\in\mathcal{C}} \frac{1}{n_+} \sum_{i\in \mathcal{I}_c^+} \{E_{j\sim P_0} \left[ \frac{e^{\frac{L(c,i,j)}{\lambda_i}}}{E_{j\sim P_0} \left[ e^{\frac{L(c,i,j)}{\lambda_i}} \right]} L(c,i,j) \right] \},
    \label{softmax_proof_3}
\end{equation}
where $P_0$ denotes uniform distribution over $\mathcal{I}_c^-$. By setting $\lambda_i=\tau$, the analogy of Eq. \eqref{softmax_proof_3} and the softmax sampling based problem (Eq. \eqref{implict_object}) is obvious. The only difference is that the index term in Eq. \eqref{softmax_proof_3} is $\frac{\ell(r_{ci}-r_{cj})}{\tau}$ but $\frac{r_{cj}-r_{ci}}{\tau}$ in Eq. \eqref{implict_object}. When choosing $\ell(t)=\log(1+\exp(-t))$, it is consistent for optimization. \par
By now, we have proven the equivalence between softmax sampling based problem (Eq. \eqref{implict_object}) and OPAUC($\beta$) objective (Eq. \eqref{object_OPAUC}). However, it is impossible to directly compute $\lambda_i$ with Eq. \eqref{softmax_proof_2}. Hence, following \cite{Faury_Tanielian_Dohmatob_Smirnova_Vasile_2020}, we propose approximation of optimal temperature parameter $\lambda_i$. A second-order Taylor expansion around 0 of Eq. \eqref{softmax_proof} yield:
\begin{equation}
\begin{aligned}
    \min_{\theta} \min_{\mathbf{\lambda}\ge 0} \frac{1}{|\mathcal{C}|} \sum_{c\in\mathcal{C}} \frac{1}{n_+} \sum_{i\in\mathcal{I}_c^+} 
    \{\lambda_i\cdot\rho + E_{j\sim P_0}[L(c,i,j)] + \\ \frac{\mathrm{Var}_j(L(c,i,j))}{2\lambda_i} + o_\infty(\frac{1}{\lambda_i})\},
\end{aligned}
\end{equation}

where $\mathrm{Var}_j(L(c,i,j))$ is defined in Eq. \eqref{var}. Solving the above equation yields approximated optimal temperature parameter $\lambda_i$:
\begin{equation}
    \lambda_i \simeq \sqrt{\frac{\mathrm{Var}_j(L(c,i,j))}{2\rho}} = \sqrt{\frac{\mathrm{Var}_j(L(c,i,j))}{-2\log\beta}}.
    \label{tau_choice}
\end{equation}
\end{proof}

\section{Proof of Theorem \ref{theorem_topk_measure}}
\begin{proof}
Suppose there are $i$ ($i<K$) positive items in Top-$K$ items of the permutation, and then we have $Recall@K=i/K$. Under this condition, easily, we can find out the case which has the maximum value of OPAUC($\beta$), where $\beta=\frac{K}{N_-}$: 
$$  \underbrace{+\cdots+}_{i} \ \underbrace{-\cdots-}_{K-i}|\underbrace{+\cdots+}_{N_+-i} \ \underbrace{-\cdots-}_{N_--K+i} $$
Hence, the maximum value of OPAUC($\beta$) is $\frac{-i^2+(N_++K)i}{N_+N_-}$. Meanwhile, since $i$ can only be integers, we derive that:
\begin{equation*}
\begin{split}
\frac{1}{N_+}\left\lfloor \frac{N_++K-\sqrt{(N_++K)^2-4N_+N_-\times OPAUC(\beta)}}{2} \right\rfloor \\ \leq 
    Recall@K. 
\end{split}
\end{equation*}
Similarly, we can find out the case which has the minimum value of OPAUC($\beta$):
$$  \underbrace{-\cdots-}_{K-i} \ \underbrace{+\cdots+}_{i}|\underbrace{-\cdots-}_{i} \ \underbrace{\cdots}_{N_++N_--K-i} $$
Hence, the minimum value of OPAUC($\beta$) is $\frac{i^2}{N_+N_-}$. Since $i$ can only be integers, we can also derive that: 
$$ Recall@K \leq \frac{1}{N_+} \left\lceil \sqrt{N_+N_-\times OPAUC(\beta)}\right\rceil. $$
These complete the proof of Eq. \eqref{proof_recall}. Noticing that for a given permutation, $Precision@K=\frac{N_+}{K}\cdot Recall@K$, where $\frac{N_+}{K}$ is a constant. Hence, we can easily derive the Eq. \eqref{proof_precision}.
\end{proof}

\section{Metrics} \label{Appendix_metrics}
 Suppose we sort the left items in descending order according to scores $r_{cj}$ for each context c. The positive item sets are denoted as $\mathcal{I}_{c,test}^+$. The detailed definitions of the widely-used metrics are summarized as follow:
\begin{itemize}
    \item Precision@$K$: metrics the fraction of positive items among the top $K$ predicted items:
    \begin{equation*}
        Precision@K = \frac{|\{j\in\mathcal{I}_{c,test}^+\ |\ Rank_j<K\}|}{K}.
    \end{equation*}
    \item Recall@$K$: metrics the fraction of all positive items that were recovered in the top K:
    \begin{equation*}
       Recall@K = \frac{|\{j\in\mathcal{I}_{c,test}^+\ | \ Rank_j<K\}|}{|\mathcal{I}_{c,test}^+|}.
    \end{equation*}
    \item NDCG@$K$ metrics the quality of recommendation through discounted importance based on position:
    \begin{equation*}
        NDCG@K = \frac{1}{\sum_{i=1}^{\min(|\mathcal{I}_{c,test}^+|,K)} \frac{1}{\log_2(i+1)}} \sum_{j\in\mathcal{I}_{c,test}^+} \frac{\mathbb{I}(Rank_j<K)}{\log_2(Rank_j+1)}.
    \end{equation*}
\end{itemize}

\begin{figure*}[t]
  \centering
  \includegraphics[width=0.8\textwidth]{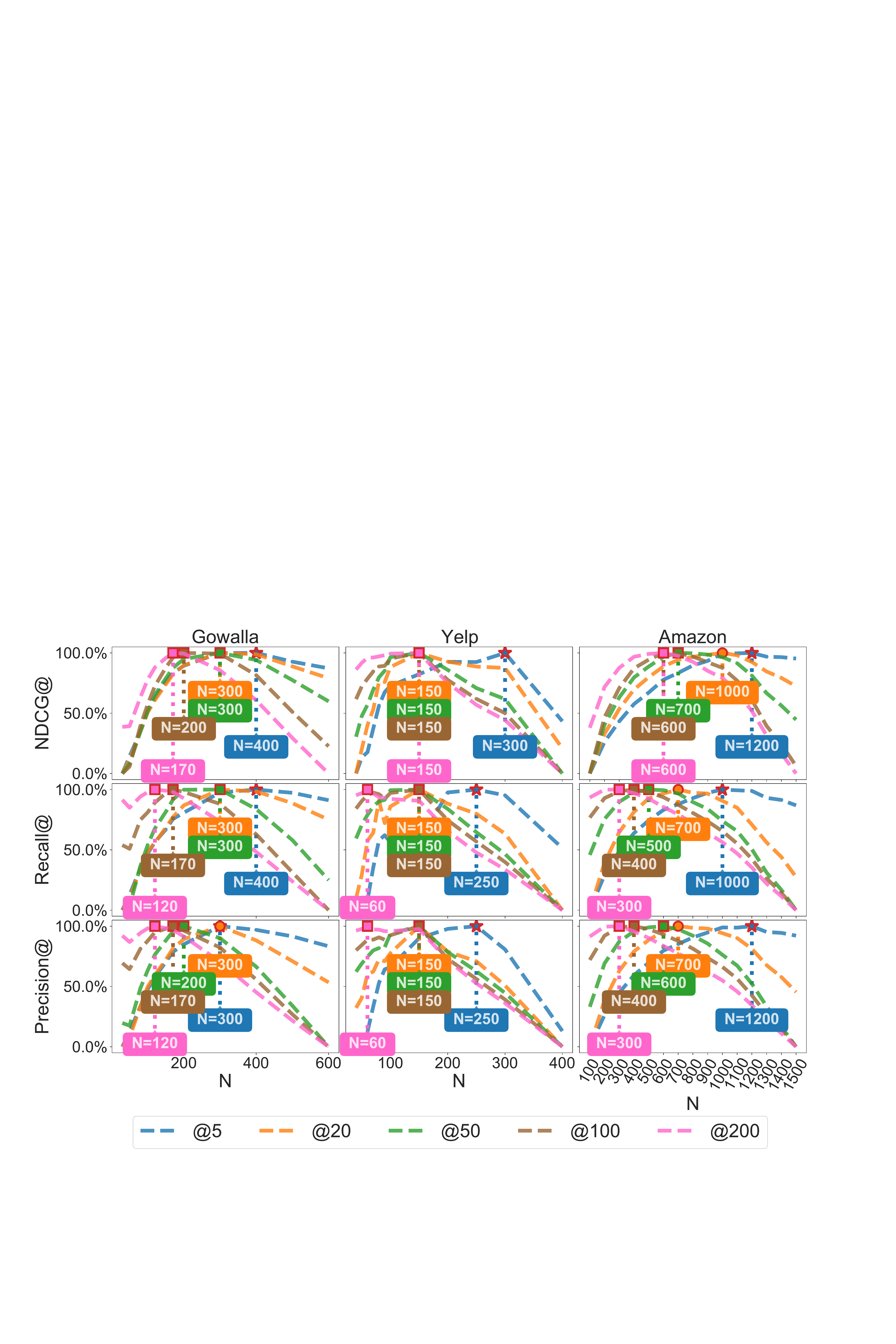}
  \caption{The effect of $N$ in DNS($M$, $N$), where $K$ is set to 5 for all three datasets.}
  \Description{For all datasets and all metrics, the lower the $K$ in Top-$K$ metrics is, the larger the $N$ in DNS($M$, $N$) when the curve achieves its maximum performance.}
  \label{fig:pool_size_appendix}
\end{figure*}

\section{Performance across a wide range of K under different sampling distribution}\label{appendix_further_K}
In this subsection, we investigate how Top-$K$ metrics across a wide range of $K$ will change under different sampling distributions. For simplicity representation, we only investigate hyperparameters $N$ in DNS($M$, $N$). As shown in Figure \ref{fig:pool_size_appendix}, we observe that: 
\begin{itemize}[leftmargin=*]
    \item The lower the $K$ in Top-$K$ metrics is, the larger the $N$ in DNS($M$, $N$) when the curve achieves its maximum performance.
    \item With a larger difference of K, there is a larger gap when the curve achieves its maximum performance.
\end{itemize}

\section{Additional Experiments With LightGCN}

\begin{table*}[t]

\caption{Performance comparison on three datasets using LightGCN. The best results are in bold and the second best are underlined.``**'' denote the improvement is significant with t-test with $p<0.05$.}
\Description{Performance comparison on three datasets using LightGCN. Our methods outperforms all baselines.}
\label{results_lightgcn}
\resizebox{.95\textwidth}{!}{%
\begin{tabular}{c|cc|cc|cc}
\hline
\multirow{2}{*}{Method} & \multicolumn{2}{c|}{Gowalla}  & \multicolumn{2}{c|}{Yelp}  & \multicolumn{2}{c}{Amazon} \\ 
\cline{2-7}
        & NDCG@50    & Recall@50       & NDCG@50    & Recall@50    & NDCG@50    & Recall@50    \\ \hline
BPR     &0.1469          & 0.2470           & 0.0742     & 0.1507         & 0.0566       & 0.1307 \\
PRIS(U) & 0.1604  & 0.2677       & 0.0831     &  0.1670        & 0.0527       & 0.1221 \\
PRIS(P) &  0.1665        &  0.2753   & 0.0870     &  0.1741        &  0.0602      &0.1369  \\
AdaSIR(U)  & 0.1804  &0.2979         &  0.0918   & 0.1808    & 0.0804   & 0.1719 \\
AdaSIR(P)  & 0.1806   & 0.2974  & 0.0914     & 0.1796         &  0.0804      & 0.1712 \\
\hline
DNS(*)  &  \underline{0.1954**}        &\underline{0.3176**}        &\underline{0.0984**}      &\underline{0.1926**}        & \underline{0.1060**}       & \underline{0.2106**} \\
Softmax-v  &\textbf{0.1991**}        &\textbf{0.3209**}   & \textbf{0.1012**}     & \textbf{0.1974**}         & \textbf{0.1100**}       &\textbf{0.2134**}  \\
\hline
\end{tabular}
}

\end{table*}

As shown in Table \ref{results_lightgcn}, we conduct additional experiments on the LightGCN model, getting similar results. Our methods DNS($M$, $N$), Softmax-v($\rho$, $N$) significantly outperform BPR and HNS baselines, which is consistent with our analysis in Subsection 6.3.

\section{Discussion}
Our analysis for BPR loss can be generalized to other loss functions, like BCE loss, Triplet loss, Softmax loss, and InfoNCE loss, which are widely applied in recommendation or other areas. Generally speaking, these loss functions have a high correlation with the AUC metric, and our conclusions also work for them. Theoretically, we have the following discussions. BCE optimizes a modified version of AUC \cite{DBLP:journals/frai/NamdarHK21}; BPR loss is a soft version of Triplet loss. Adjusting margin term in Triplet loss is equal to adjusting M in DNS($M$, $N$); Softmax-v($\rho$, $N$) is the upper bound of Softmax loss and InfoNCE loss. Adjusting the temperature in Softmax loss and InfoNCE loss is equal to adjusting $\rho$ in Softmax-v($\rho$, $N$).

\end{document}